\documentstyle[12pt,dina4p,epsfig]{article}
\newcommand{\be}{\begin{eqnarray}}
\newcommand{\ee}{\end{eqnarray}}
\newcommand{\nn}{\nonumber}
\newcommand{\eps}{\epsilon}
\renewcommand{\d}{\mbox{d}}
\newcommand{\dis}{\displaystyle}
\newcommand{\g}{\gamma}
\newcommand{\G}{\Gamma}
\title{
  \vskip-2cm
  {\baselineskip16pt
    \centerline{\normalsize \tt DESY 96-035 \hfill ISSN 0418-9833}
    \centerline{\normalsize \tt hep-ph/9602418 \hfill}
    \centerline{\normalsize \tt Februar 1996 \hfill}
  }
  \vskip2cm
  {\bf 
    Inclusive One-- and Two--Jet Cross Sections
    in $\g\g$ Reactions at $e^+e^-$ Colliders 
  }
  \author{
    {T. Kleinwort,  G. Kramer} \\
    {II. Institut f\"ur Theoretische Physik}\thanks
     {Supported by Bundesministerium f\"ur Forschung und
     Technologie, Bonn, Germany under Contract 05\,6HH93P(5) and
     EEC Program ``Human Capital and Mobility'' through Network
     ``Physics at High Energy Colliders'' under Contract
     CHRX--CT93--0357 (DG12 COMA) }
    \\
    {Universit\"at Hamburg} \\
    {D - 22761 Hamburg, Germany}
  }
  \date{}
}
\begin{document}
\maketitle
\vspace{3cm}
\begin{abstract}
\thispagestyle{empty}
We have calculated inclusive one-- and two--jet production in 
photon--photon collisions superimposing direct, single resolved and
double resolved cross sections for center of mass energies of the
LEP1, LEP2 and NLC range. The direct and single resolved cross
sections are calculated up to next--to--leading order. The double
resolved two--jet cross section is calculated only in LO with a $k$
factor estimated from the NLO one--jet cross section. Various
differential cross sections as functions of transverse momenta and
rapidities of the jets are evaluated.
\end{abstract}
\newpage
\setcounter{page}{1}
\section{Introduction}

In a recent paper we presented theoretical results for the
production of high--$p_T$ jets in almost--real photon--photon
collisions \cite{KleKra} and compared them with recent experimental data
for inclusive one-- and two--jet cross sections from the 
two TRISTAN collaborations, TOPAZ \cite{TOPAZ} and AMY \cite{AMY}.
Due to the increased c.m.\ energy of the TRISTAN ring as compared to earlier
low energy experiments at PETRA and PEP the TOPAZ and AMY cross sections
extend to transverse momenta of up to 8 $GeV$. It is hoped that coming
data from $\g\g$ collisions at LEP1 and particularly from LEP2 
will enlarge the $p_T$ range further.
Production of high--$p_T$ jets probes the short--distance dynamics of
photon--photon reactions. In addition to providing tests of perturbative QCD, 
data from $\g\g$ reactions give us information on the 
photon structure function.

In leading order QCD (LO) three distinct classes of contributions to the
cross section \cite{DreGod} are identified: (i) The direct
contribution, in which the two photons couple directly to quarks
and neither of the photons is resolved into its partonic constituents
(DD component in the following). (ii) The single--resolved contribution,
where one of the photons interacts with the partonic constituents of the
other (DR component). (iii) The double--resolved contribution, where
both photons are resolved into partonic constituents before the hard
scattering process takes place (RR component). In the DD component
we have only the two high--$p_T$ jets in the final state and no additional
spectator jets. In the DR case one spectator jet coming from low transverse
momentum fragments of one of the photons is present and in the
RR component we have two such spectator or photon remnant jets. 
Experimental separation of the three classes is in principle
possible and depends whether the experimental arrangements allow
to tag the spectator jets.

In next--to--leading order QCD (NLO) the photon--quark collinear singularity
arising in the DD and DR components is subtracted and absorbed into the
photon structure function in accord with the factorization theorem.
This subtraction procedure at the factorization scale $M$ introduces an
interdependence of the three components so that a unique separation into 
DD, DR and RR classes is not possible anymore. This means that in NLO
all three components must be considered together and consistently be 
calculated up to NLO in the photon structure functions \cite{GRV,GSACFG} 
and in the hard scattering cross sections.

The photon structure functions are inherently non-perturbative quantities
whose magnitude and dependence on the fractional momentum $x$ of the outgoing
parton must be measured at a reference scale ${M_0}^2$. The change with
$M^2$ is obtained from perturbative QCD evolution equations.

Complete NLO calculations for high--$p_T$ jet production in
$\g\g$ reactions have been done previously for the inclusive 
single jet cross section \cite{KleKra,Aetal} and compared to experimental
data from TRISTAN \cite{TOPAZ,AMY}. The double resolved contribution
to the NLO single jet inclusive cross section has been investigated
in \cite{Gord} for TRISTAN, LEP1 and LEP2 energies in order to study the
dependence on the photon structure function. The restriction to the RR
contribution was motivated in \cite{Gord} that by tagging the two spectator
jets it could be isolated experimentally from the other components.
Unfortunately this is possible only in LO since NLO corrections to the
DD and DR components can produce spectator jets due to the separation
of singularities with the help of kinematic constraints.

In our previous work \cite{KleKra} we also calculated the complete NLO
inclusive dijet cross sections for the DD and DR contributions and 
estimated the RR contribution by a LO calculation with $k$ factors 
taken from the NLO inclusive single--jet cross section. This was 
justified since at TRISTAN energies, for which the calculation was done,
the RR cross section contributes only a small fraction of the total
sum in the high--$p_T$ region. In this paper we extend these calculations
to LEP1, LEP2 and NLC collider energies.
Although at these higher energies and at moderate $p_T$ the double
resolved cross section is much more important, we expect that the estimate 
of the RR dijet cross sections with a $k$ factor  will still
be reliable. In addition we shall
present complete NLO predictions for the single--jet inclusive 
cross section at these energies which we expect 
to be measured first. Since in our earlier work where we compared also 
to experimental dijet cross sections measured by TOPAZ and AMY no
details for the calculation of the cross sections were given, 
we shall present them in this longer paper.

The outline of the rest of the paper is as follows. In section 2 we first
give the DD cross section in  leading order also to fix the notation.
Then we present the details of the next--to--leading order calculation
for the DD two--jet cross section in section 3.

The calculation of the DR and RR cross sections is based on the work 
of \cite{KlaKra} and \cite{KS}. In \cite{KlaKra} the formalism for
inclusive one-- and two--jet cross sections is worked out and applied
to the calculation of the direct process on jet production in 
low $Q^2$ $ep$ collisions. In \cite{KS} the inclusive
one--jet cross section for $ep$ collisions with a resolved photon was 
calculated which can easily be applied to the RR cross section in 
$\g\g$ reactions.

The numerical results for the DD, DR and RR one-- and two--jet cross 
sections are given in section 4. In section 5 we summarize our results
and draw some conclusions.

\section{Leading Order Cross Section}

\subsection{Photon Spectrum}

The cross sections which we have computed are for kinematical conditions 
as we expect them for $\g\g$ collisions at LEP1, LEP2 and the NLC.
For LEP1 and LEP2 the
spectrum of the virtual photons is described in the Weizs\"acker--Williams
approximation (WWA) by the formula \cite{FMNR}
\be
   F_{\g/e}(x_a) &=& \frac{\alpha}{2\pi}\Bigg\{
   \frac{1+(1-x_a)^2}{x_a} \ln{\frac{E^2\Theta_c^2(1-x_a)^2+m_e^2x_a^2}
                                     {m_e^2x_a^2}} \nn\\
   & & +2(1-x_a)\left[\frac{m_e^2 x_a}{E^2\Theta_c^2
        (1-x_a)^2+m_e^2x_a^2}-\frac{1}{x_a}\right]\Bigg\}\label{f1},
\ee
where $m_e$ is the electron mass and $x_a=(q_ak_b)/(k_bk_a)\simeq
E_\g/E$ is the fraction of the initial positron energy transferred
to the photon with $E$ being the beam energy of the incoming positrons
and electrons, respectively.
$k_a$($k_b$) is the four momentum of the incoming positron
(electron) and $q_a=k_a-k_a'$ is the four momentum of the virtual photon
emitted by the positron. The momentum fraction on the electron side
is denoted as $x_b$ with $x_b=(k_aq_b)/(k_ak_b)$ and $q_b=k_b-k_b'$. 

In the equivalent photon approximation, the cross section for
$e^++e^-\to e^++e^-+X$ with arbitrary final state $X$ is then given 
by the convolution
\be
   \d\sigma\left(e^++e^-\to e^++e^-+X\right)=\int\limits_{x_{a,\min}}^1\!\!\!
   \d x_a \!\!\!\int\limits_{x_{b,\min}}^1 \!\!\!\d x_b\,
   F_{\g_a/e}(x_a)F_{\g_b/e}(x_b)\, 
   \d\sigma(\g_a\g_b\to X)\label{f2},
\ee
where $\d\sigma(\g_a\g_b\to X)$ denotes the cross section for
$\g_a\g_b\to X$ with real photons of energies $E_{\g_a} = x_a E$
and $E_{\g_b} = x_b E$, respectively.

The LEP1 cross sections are calculated with $\sqrt{S} = 2E = 90\, GeV$.
The angle $\Theta_c$ is the maximum angle under which the positrons
(electrons) are tagged, which we assume $\Theta_c = 3.0^\circ$. 
For the LEP2 cross 
section we have chosen $\sqrt{S} = 175\, GeV$ and $\Theta_c = 30\, mrad$.
For the NLC collider option the photon spectra will be specified later
when we present our results. 

\subsection{Jet Cross Sections in LO}

In this section we write down the leading order cross section for the
production of jets with two direct photons, i.e. the DD cross section
in LO. This is given by the differential cross section for two--jet
production in the quark parton model, i.e. the cross section for 
$\g_a\g_b\to q\bar{q}$, which must be convoluted with the
photon distribution functions according to (\ref{f2}). 

The final state quark (antiquark) has momentum $p_1$($p_2$) which can
be expressed by their transverse momentum $p_T$ and rapidities $\eta_1$
and $\eta_2$. The convention is that the $z$ direction is parallel to
the electron beam direction. From energy and momentum conservation one
obtains 
\be
   x_a &=& \frac{p_T}{2E}\left(e^{-\eta_1}+e^{-\eta_2}\right) \label{f3}\\
   x_b &=& \frac{p_T}{2E}\left(e^{ \eta_1}+e^{ \eta_2}\right) \label{f4}.
\ee
Thus, the kinematical variables of the two jets are related to the
scaling variables $x_a$ and $x_b$. Under the actual experimental
conditions $x_a$ and $x_b$ are restricted to fixed intervals,
$x_{\min}<x_a, x_b<x_{\max}<1$. We shall disregard these constraints
and allow $x_a$ and $x_b$ to vary in the kinematically allowed range
$x_{a,b,\min}<x_a,x_b<1$, where
\be
   x_{a,\min} = \frac{p_T e^{-\eta_1}}{2E-p_Te^{\eta_1}} \label{f5}
\ee
and a similar equation for $x_{b,\min}$. From eq. (\ref{f3}) and
(\ref{f4}), we can express $x_b$ as a function of $p_T$, $\eta_1$ and
$x_a$: 
\be
   x_b &=& \frac{x_ap_Te^{\eta_1}}{2x_aE-p_Te^{-\eta_1}}\label{f6}.
\ee
Depending on $x_a$, different regions of the photon energy $x_b$
contribute. Whereas for fixed $x_a$ and $x_b$, i.e. fixed energies for
both photons, $\eta_1$ and $\eta_2$ are fully determined by $p_T$ (up
to sign ambiguities), $\eta_1$ and $\eta_2$ are allowed to vary due to
the kinematical range of the photon energies in the intervals
$x_{a,b}\in [x_{a,b,\min},1]$. $x_{a,b,\min}$ are obtained from (5)
and are equal to ${p_T}^2/E^2$. 

The two--jet cross section for $e^++e^-\to e^++e^-+\mbox{jet}_1 +
\mbox{jet}_2$ is obtained from
\be
   \frac{\d^3\sigma}{\d p_T\d\eta_1\d\eta_2} &=&
   x_aF_{\g_a/e}(x_a)x_bF_{\g_b/e}(x_b)\frac{\d\sigma}{dt}
   \left(\g_a\g_b\to p_1p_2\right)\label{f7}.
\ee
$\dis\frac{\d\sigma}{dt}$ stands for the differential cross section of the
process $\g_a\g_b\to p_1p_2$. The invariants of this process
are $s=(q_a+q_b)^2$, $t=(q_b-p_1)^2$ and $u=(q_b-p_2)^2$. They are
expressed by the final state variables $p_T$, $\eta_1$ and $\eta_2$
and the initial state momentum fractions $x_a$ and $x_b$:
\be
   s &=& 4x_ax_bE^2 \label{f8} \\
   t &=& -2x_aEp_Te^{\eta_2} = -2x_bEp_Te^{-\eta_1} \label{f9} \\
   u &=& -2x_aEp_Te^{\eta_1} = -2x_bEp_Te^{-\eta_2} \label{f10}.
\ee
So, the dependence of the two--jet cross section on $p_T$, $\eta_1$
and $\eta_2$ is determined through the two photon distribution
functions
and the cross section for the $\g\g$ subprocess, which depends
on $s$, $t$ and $u$.

For the inclusive one--jet cross section, we must integrate over one
of the rapidities in (\ref{f7}). We integrate over $\eta_2$ and
transform to the variable $x_a$ using (\ref{f3}). The result is the cross
section for $e^++e^-\to e^++e^-+\mbox{jet}+X$, which depends on $p_T$
and $\eta$:
\be
   \frac{\d^2\sigma}{\d p_T\d\eta}&=&
   \int\limits_{x_{a\min}}^1 \d x_a x_a F_{\g_a/e}(x_a)
   F_{\g_b/e}(x_b) \frac{4Ep_T}{2x_aE-p_Te^{-\eta}}
   \frac{\d\sigma}{\d t}\left(\g_a\g_b\to p_1p_2\right)\label{f11}.
\ee
Here, $x_b$ is given by (\ref{f6}) with $\eta_1=\eta$. 

The cross section for the subprocess $\g_a\g_b\to
q_i\bar{q_i}$ is well known and is given by
\be
   \frac{\d\sigma}{\d t}\left(\g_a\g_b\to q_i\bar{q_i}\right) &=&
   \frac{2\pi\alpha^2}{s^2}N_CQ_i^4
   \left(\frac{u}{t}+\frac{t}{u}\right)\label{f12}
\ee 
The index $i$ denotes the quark flavour and $Q_i$ the quark charge.
$N_C$ is the number of colours. All quarks are considered massless. 

\section{ Next--To--Leading Order Cross Sections}

The next--to--leading order corrections are calculated with the help
of dimensional regularization.  In the following subsections we shall
consider the virtual and the real corrections needed to obtain a
finite cross section for the DD case in the limit $n$ $\to$ 4.

\subsection{Virtual Corrections up to O($\alpha^2\alpha_s$)}

The one loop diagrams for $\g\g\to q\bar{q}$ have an additional virtual
gluon, which leads to an extra factor $\alpha_s$. These diagrams must
be multiplied with the LO diagrams to produce the virtual corrections
to the 2 $\to$ 2 cross section up to O($\alpha^2\alpha_s$). These
corrections are well known for many years now \cite{BKGK,ABDFS}.
The result can also be obtained from the virtual correction for
$\g q\to g q$ \cite{KlaKra} by taking only the diagrams without the three
gluon vertex and substituting the appropriate colour factors. The result
can be written as
\be
   H_V(\g\g\to q_i\bar{q_i}) &=& e^4Q_i^4 \mu^{4\eps}
   \left(\frac{4\pi\mu^2}{s}\right)^\eps 
   \frac{\alpha_s}{2\pi}
   \frac{\G(1-\eps)}{\G(1-2\eps)}  
   2N_CC_FV_\g(s,t,u) + O(\eps).
\ee
$H_V$ gives the virtual correction to the corresponding reactions up 
to the $n$--dimensional phase space factor
\be
   \frac{\mbox{dPS}^{(2)}}{\d t} &=& \frac{1}{\Gamma(1-\eps)}
   \left(\frac{4\pi s}{ut}\right)^\eps \frac{1}{8\pi s}
\ee
and the flux factor $1/(2s)$.
The expression for $V_\g(s,t,u)$ can be found in appendix
A. The singular terms $\propto$ $1/\eps^2$ and $1/\eps$ are proportional to 
the LO cross section
\be
   T_\g^{(\eps)} &=& (1-\eps)\left[\left(\frac{t}{u}+\frac{u}{t}\right)
            (1-\eps)-2\eps\right]
\ee
with $\eps=(4-n)/2$.
\subsection{Real Corrections up to O($\alpha^2\alpha_s$)}

\begin{figure}[ht]
 \begin{center}
  \begin{picture}(3,3)
    \epsfig{file=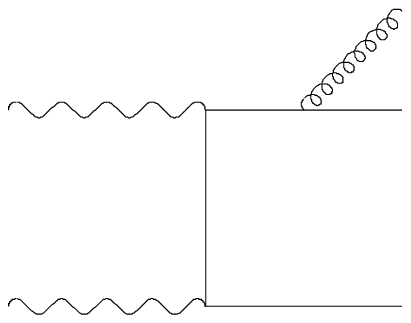,bbllx=240pt,bblly=360pt, bburx=360pt,bbury=480pt,%
            height=3cm}
  \end{picture}
\hphantom{xxxx}
  \begin{picture}(3,3)
    \epsfig{file=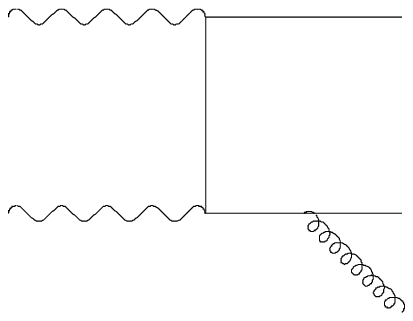,bbllx=240pt,bblly=360pt, bburx=360pt,bbury=480pt,%
            height=3cm}
  \end{picture}
\hphantom{xxxx}
  \begin{picture}(3,3)
    \epsfig{file=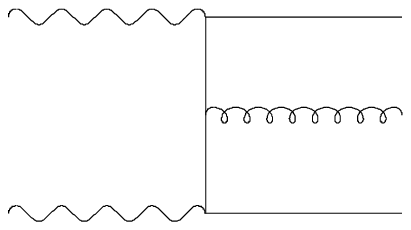,bbllx=240pt,bblly=360pt, bburx=360pt,bbury=480pt,%
            height=3cm}
  \end{picture}\par
 \end{center}
 \caption{\label{fig1}}
\end{figure}

For the 2 $\to$ 3 contributions up to O($\alpha^2\alpha_s$),
 we have to take into account all diagrams with an additional gluon in
the final state. The diagrams are shown in Fig.~\ref{fig1}.

The four--vectors of these subprocesses will be labeled by
$q_aq_b\to p_1p_2p_3$. The invariants will be denoted by
$s_{ij}=(p_i+p_j)^2$, $(i,j=a,b,1,2,3)$. For massless partons, the 2 $\to$ 3
contributions contain singularities at $s_{ij}=0$. They can be
extracted with the dimensional regularization method and can be cancelled
against those which originate from the one--loop contributions or are 
absorbed in the renormalized photon structure functions. 

\begin{figure}[ht]
 \begin{center}
  \begin{picture}(4,3)
    \epsfig{file=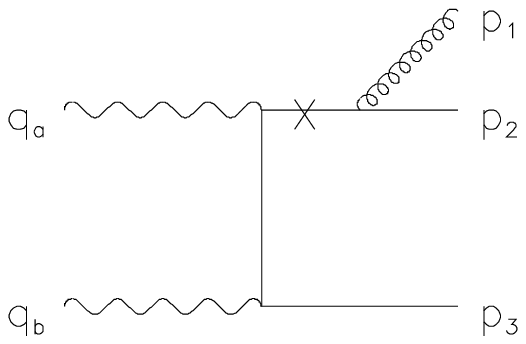,bbllx=240pt,bblly=360pt, bburx=375pt,bbury=480pt,%
            height=3cm}
  \end{picture}
\hphantom{xxxx}
  \begin{picture}(4,3)
    \epsfig{file=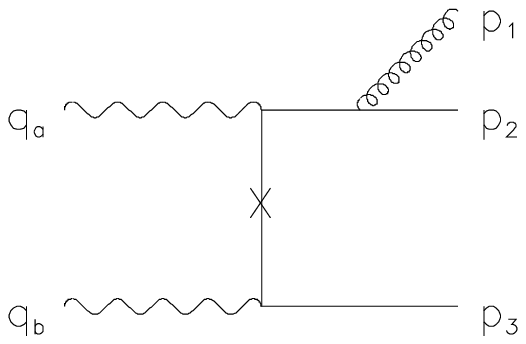,bbllx=240pt,bblly=360pt, bburx=375pt,bbury=480pt,%
            height=3cm}
  \end{picture}\par
  \caption{\label{fig2}}
 \end{center}
\end{figure}

To achieve this, we go through the same steps as described for example in
\cite{KlaKra}. First, we calculated the 2 $\to$ 3 subprocesses in 
$n$ dimensions. In our case we have two classes of singularities.
Examples of these are shown in Fig.~\ref{fig2}. 
The X marks the propagator
leading to the divergence. In the first graph, the vanishing of the
invariant $s_{12}$
leads to a final state singularity and the second graph becomes
singular for $s_{b3}=0$, which then leads to an initial state
singularity. 

When squaring the sum of all diagrams in Fig.~\ref{fig2}, 
we encounter terms
where more than one of the invariants become singular, e.g. when the
gluon momentum $p_1\to0$, so that $s_{12}=0$ and $s_{13}=0$. These
infrared singularities are disentangled by a partial fractioning
decomposition, so that every term has only one vanishing denominator.
Non--singular terms are discarded.
It turns out that in this limit 
the results are always proportional to the LO cross
sections involved in the hard scattering, namely $T_\g$ and $T_q$,
where $T_q$ stands for $\g q\to gq$, so that they can be written as
\be
   H_{F,I} &=& K_{F,I} T_{\g,q}.
\ee
Here $F$ and $I$ denote the contributions originating from 
the final(F) and initial(I) state singularities, respectively.

The last step is to integrate the decomposed matrix elements
analytically in the region $s_{ij}\le ys$, where $y$ is the upper bound for
the integration and has to be chosen small enough 
so that it is justified to neglect terms of $O(y)$, which was assumed 
already in the partial fractioning above.
This means we use an invariant mass cut to separate the genuine
$2\to3$ contributions from the $2\to2$ contributions. For the terms
with final state singularities $y$ determines the phase space region
where two partons, i.e.\ a quark and a gluon, are combined in one jet.
In the case of initial state singularities $y$ determines the boundary
between the remnant jet of the photon and the $2\to2$ hard scattering
$\g q\to gq$. In \cite{BKGK} the separation of the final state 
singularities was done with the  
Sterman--Weinberg parameters $\varepsilon$ and $\delta$,
where $\varepsilon$ is the cut on the gluon energy and
$\delta$ measures the angle between the momenta of the quark and the gluon
in the recombination of a quark and a gluon into one jet.
The integration produces terms $\propto 1/\eps^2$ and
$1/\eps$, which cancel against those in the virtual corrections or be
absorbed into the photon structure functions. In the following we
shall give the results for the final and the initial state
singularities separately.

\subsubsection{Final State Singularities}

\begin{figure}[hb]
 \begin{center}
  \begin{picture}(5,5)
    \epsfig{file=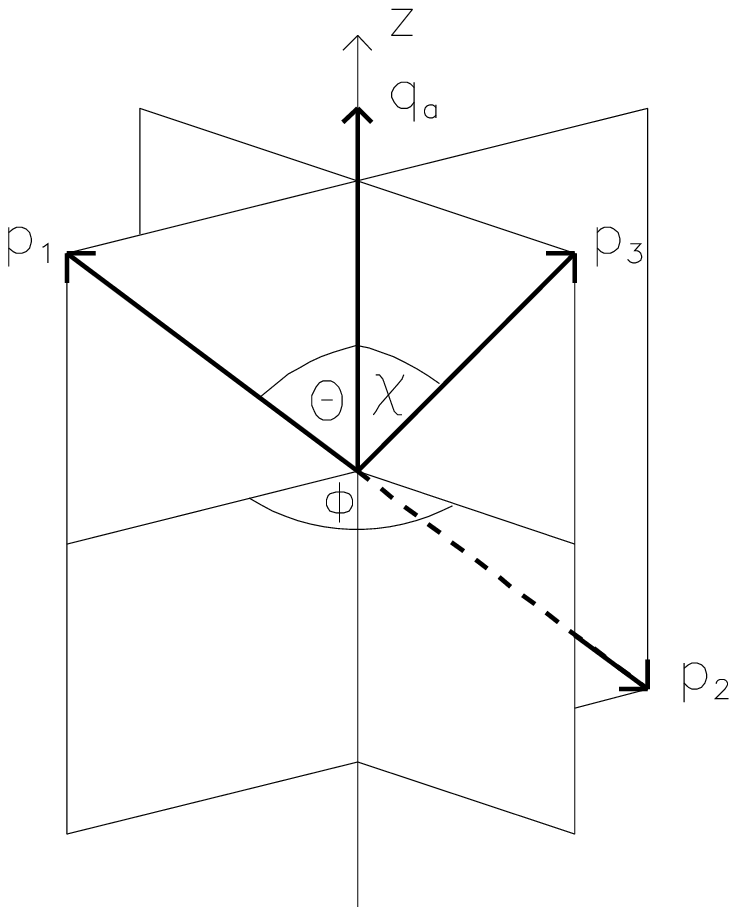,bbllx=195pt,bblly=295pt, bburx=405pt,bbury=555pt,%
            height=5cm}
  \end{picture}
  \caption{\label{fig4}}
 \end{center}
\end{figure}

In this subsection, we assume that after partial fractioning the 2 $\to$
3 matrix elements are singular only for $s_{12}=0$. For the
integration over the singular phase space region we choose as
coordinate system the c.m.\ system of the partons $p_1$ and $p_2$.
The angles of the other parton three--momenta $q_a$ and $p_3$ with
respect to $p_1$ and $p_2$ are shown in Fig.~\ref{fig4}. $\chi$ is the
angle between the momenta $q_a$ and $p_3$, $\theta$ is the 
angle between $q_a$ and $p_1$, and $\phi$ is the azimuthal angle
between the planes defined by $q_a$ and $p_1$ and $p_a$ and $p_3$, 
respectively. Instead of $\theta$ we also use the variable 
\be
   b &=& \frac12(1-\cos\theta).
\ee
The angle $\phi$ can be integrated out easily, because the
matrix elements do not depend on it in the limit that non--singular
terms are discarded.

Here, we define the invariants
\be
   s &=& (q_a+q_b)^2 \\
   t &=& (q_b-p_1-p_2)^2-2p_1p_2 \\
   u &=& (q_b-p_3)^2
\ee
which differ from the corresponding two-body invariants, but are equal to 
them for $p_1=0$ or $p_2$ collinear with $p_1$. 
The variable to be integrated is
\be
   z'&=&\frac{p_1.p_2}{q_a.q_b}.
\ee
The three--body phase space in $n$ dimensions can be factorized into
\be
   \mbox{dPS}^{(3)}=\mbox{dPS}^{(2)}\mbox{dPS}^{(r)},
\ee
where
\be
   \frac{\mbox{dPS}^{(2)}}{\d t}&=&\frac{1}{\G(1-\eps)}
   \left(\frac{4\pi s}{tu}\right)^\eps\frac{1}{8\pi s} \label{fdp}
\ee
and
\be
   \mbox{dPS}^{(r)}&=&\left(\frac{4\pi}{s}\right)^\eps
   \frac{\G(1-\eps)}{\G(1-2\eps)}\frac{s}{16\pi^2}\frac{1}{1-2\eps}
   \d\mu_F
\ee
with
\be
   \d\mu_F &=& \d z'z'^{-\eps}\left(1+\frac{z's}{t}\right)^{-\eps}
   \frac{\d b}{N_b}b^{-\eps}(1-b)^{-\eps}\frac{\d\phi}{N_\phi}
    \sin^{-2\eps}\!\phi.
\ee
$N_b$ and $N_\phi$ are normalization factors:
\be
   N_b &=& \int\limits_{0}^{1}\d b\, b^{-\eps}(1-b)^{-\eps} = 
   \frac{\G^2(1-\eps)}{\G(2-2\eps)} \\
   N_\phi&=& \int\limits_{0}^{\pi}\d\phi\sin^{-2\eps}\!\phi =
   4^\eps\pi\frac{\G(1-2\eps)}{\G^2(1-\eps)}.
\ee
The full range of integration is given by $z'\in[0,-t/s]$, $b\in[0,1]$
and $\phi\in[0,\pi]$. The singular region is defined by the requirement
that partons $p_1$ and $p_2$ are recombined, which means $s_{12}\to 0$. 
We integrate over this region analytically up to $s_{12}\le ys$, which
restricts the range of integration to $0\le z'\le\min\{-t/s,y\}\equiv
y_F$. 

In the $\g\g$ case we obtain the final state matrix element 
in the following form
\be
   \int\mbox{dPS}^{(r)}H_F&=&
   Q_i^4\mu^{4\eps}\left(\frac{4\pi\mu^2}{s}\right)^\eps
   \frac{\alpha_s}{2\pi}\frac{\G(1-\eps)}{\G(1-2\eps)}2N_CC_F
   F_\g(s,t,u)+O(\eps).
\ee
The factor $F_\g(s,t,u)$ can be found in Appendix B. It contains
infrared and collinear singularities, which cancel against those in
the virtual corrections.

\subsubsection{Initial State Singularities}

Here we integrate over the singularity $s_{b3} = 0$, we use the same 
$p_1$--$p_2$ c.m.\ system as in the last section. 
In this case, the outgoing quark $p_3$ is collinear to the incoming
photon momentum $q_b$. So it becomes part of the photon remnant
and $p_1$ and $p_2$ are the momenta of the final state in
the 2--body subprocess. We introduce the new variable
\be
   z_b &=& \frac{p_1p_2}{q_aq_b}\in \left[X_b,1\right],
\ee
where $X_b=(p_1p_2)/(q_ak_b)\simeq E_{q}/E$ 
is the fraction of the initial electron energy
transferred to the quark.

We have now the following definition for the Mandelstam variables
\be
   s &=& (q_a+z_bq_b)^2\\
   t &=& (q_a-p_1)^2\\
   u &=& (q_a-p_2)^2
\ee
and the variable, which parametrizes the singular region, is now
\be
   z''&=&\frac{q_bp_3}{q_aq_b}.
\ee
Again the three--body phase space factorizes into
\be
   \mbox{dPS}^{(3)}=\mbox{dPS}^{(2)}\mbox{dPS}^{(r)},
\ee   
where $\mbox{dPS}^{(2)}$ is again the phase space of the 2 $\to$ 2
process now in the limit $q_bp_3\to0$ given by (\ref{fdp}), 
and 
\be
   \mbox{dPS}^{(r)}&=&\left(\frac{4\pi}{s}\right)^\eps
   \frac{\G(1-\eps)}{\G(1-2\eps)}\frac{s}{16\pi^2}H_b(z'')
   \d\mu_I,
\ee
where
\be
   \d\mu_I &=& \d z''z''^{-\eps}\frac{\d z_b}{z_b}
   \left(\frac{z_b}{1-z_b}\right)^\eps
   \frac{\d\phi}{N_\phi}\sin^{-2\eps}\!\phi
   \frac{\G(1-2\eps)}{\G^2(1-\eps)}
\ee
and
\be
   H_a(z'') &=& \left(1+\frac{z''}{z_a}\right)^{-1+2\eps}
     \left(1-\frac{z''}{1-z_a}\right)^{-\eps} = 1+O(z'')
\ee
can be approximated by 1, because it leads only to negligible terms of
$O(y)$. The full region of integration is given by $z''\in[0,-u/s]$,
$z_b\in[X_b,1]$ and $\phi\in[0,\pi]$, where again the dependence on
$\phi$ can easily be integrated out. 
The singular region where the integration is
done analytically is given by the requirement $0\le z''\le
\min\{-u/s,y\}\equiv y_I$.

The result is
\be
   \int\mbox{dPS}^{(r)}H_{I}&=&\int\limits_{X_b}^{1}\frac{\d z_b}{z_b}
   Q_i^4\mu^{4\eps}\left(\frac{4\pi\mu^2}{s}\right)^\eps\frac{\alpha_s}{2\pi}
   \frac{\G(1-\eps)}{\G(1-2\eps)} N_CC_FI_\g(z_b;s,t,u) + O(\eps).
\ee
Again, $I_\g(z;s,t,u)$ can be found in the appendix.

In the case of initial state singularities the following quark-photon
matrix element factorizes
\be
   T_q^{(\eps)}(s,t,u) &=& (1-\eps)\left[\left(-\frac{t}{s}-\frac{s}{t}\right)
                  (1-\eps)+2\eps\right].
\ee
The case $z''=q_ap_3/q_aq_b\to0$ leads to the same result with 
$(z_b\leftrightarrow z_a)$.
The term $I_\g$ shows explicitly the pole in $1/\eps$ proportional
to 
\be
   P_{q\leftarrow\g}(z)&=&N_CQ_i^2\left[2z^2-2z+1\right].
\ee
This function appears in the evolution equation of the photon
structure function as an inhomogeneous or so--called point--like term.
Therefore, the photon initial state singularities can be absorbed
into the photon structure function.  

\subsection{DD Jet Cross Section in NLO}

To obtain a finite cross section for the DD case, we must add the
parts considered in section 3.1 and 3.2. Then the poles in $1/\eps$
and $1/\eps^2$ cancel and we can take the limit $\eps\to0$. 
The result is a special kind of two--jet cross section, where the
recombination of two partons into one jet or the recombination of a
parton with the photon remnant jet is done with an invariant mass
cut-off $y$. Including the NLO corrections, we get
\be
   \frac{\d^3\sigma}{\d p_T\d\eta_1\d\eta_2}&=&
   x_aF_{\g_a/e}(x_a)x_bF_{\g_b/e}(x_b)\Bigg[
   \frac{\d\sigma}{\d t}(\g_a\g_b\to p_1p_2)\nn\\
   && + \frac{\d\tilde{\sigma}}{\d t}(\g_a q\to p_1p_2)
      + \frac{\d\tilde{\sigma}}{\d t}(q\g_b\to p_1p_2)\Bigg]\label{f70}.
\ee
In (\ref{f70}), $\dis\frac{\d\sigma}{\d t}(\g_a\g_b\to p_1p_2)$ stands
for the two--body contribution in LO and NLO together with
analytically integrated contributions of the soft and collinear
divergent regions of the three--parton final state. The contributions
from the initial state singularities are denoted 
$\dis\frac{\d\tilde{\sigma}}{\d t}(\g_a q\to p_1p_2)$ and 
$\dis\frac{\d\tilde{\sigma}}{\d t}(q\g_b\to p_1p_2)$, where either the
photon $\g_b$ or $\g_a$ has collinear singular terms removed,
respectively. 

The two--body contribution can be written as
\be
   \frac{\d\sigma}{\d t}(\g_a\g_b\to p_1p_2) &=& C
   T_\g+\frac{\alpha_s^2(\mu^2)}{2\pi}C(T_\g A_\g+B_\g),
\ee
where
\be
   A_\g&=&C_F\left[\frac{2}{3}\pi^2+\ln^2\frac{-t}{s}+\ln^2\frac{-u}{s}
   -2\ln^2y_F-3\ln y_F\right]\\
   B_\g&=&C_F\Bigg[2\ln\frac{-t}{s}+2\ln\frac{-u}{s}
   +3\frac{u}{t}\ln\frac{-t}{s} \
   +3\frac{t}{u}\ln\frac{-u}{s}\nn\\
   &&+\left(2+\frac{u}{t}\right)\ln^2\frac{-u}{s}
   +\left(2+\frac{t}{u}\right)\ln^2\frac{-t}{s}\Bigg],\\
   T_\g&=& \left(\frac{u}{t}+\frac{t}{u}\right), \mbox{ and }
   C=\frac{2\pi\alpha^2}{s^2}Q_i^4N_C.
\ee

The cut dependence of the final state corrections in $A_\g$ is
contained in the $y_F$ dependent terms. For $y_F\to0$, these terms
behave like ($-\ln^2y_F$), which leads to unphysical negative cross
sections for very small $y_F$. Thus if $y$ is used as a physical cut,
it must be sufficiently large. In most applications, we shall use
these results for computing inclusive cross sections, in which the $y$
dependence of the two--jet cross section cancels against the $y$
dependence of the numerically calculated three--jet cross section.

The two--jet cross section for the initial state is calculated from
\be
   \frac{\d\tilde{\sigma}}{\d t}(q\g_b\to p_1p_2) &=& 
   \frac{\d\hat{\sigma}}{\d t}(q\g_b\to p_1p_2)\frac{\alpha}{2\pi}\nn\\
   &&\int\limits_{X_a}^1\frac{\d z_a}{z_a}
   \left[P_{q\leftarrow\g}(z_a)\left(\ln\left(\frac{(1-z_a)y_Is}
   {z_aM_a^2}\right)-1\right)+N_CQ_i^2\right]\label{f79a}\\
   \frac{\d\tilde{\sigma}}{\d t}(\g_a q\to p_1p_2) &=& 
   \frac{\d\hat{\sigma}}{\d t}(\g_a q\to p_1p_2)\frac{\alpha}{2\pi}\nn\\
   &&\int\limits_{X_b}^1\frac{\d z_b}{z_b}
   \left[P_{q\leftarrow\g}(z_b)\left(\ln\left(\frac{(1-z_b)y_Is}
   {z_bM_b^2}\right)-1\right)+N_CQ_i^2\right].\label{f79b}
\ee
In (\ref{f79a}) and (\ref{f79b}), $M_a$ resp.~$M_b$ are the
factorization scales. The dependence on $M_a$ resp.\ $M_b$ 
 must cancel against the $M_a$ resp.\ $M_b$ 
dependences of the LO DR and RD contributions.
The cross section $\dis\frac{\d\hat{\sigma}}{\d t}$ has the following form
\be
   \frac{\d\hat{\sigma}}{\d t}(\g q\to qg) &=&
   \frac{2\pi\alpha\alpha_s}{s^2}Q_i^2C_F\left(-\frac{s}{t}-\frac{t}{s}\right).
\ee
All the results of this section are for the $\overline{\mbox{MS}}$
subtraction and renormalization scheme.
\section{Inclusive One-- and Two--Jet Cross Sections}
In this section we present some characteristic numerical results for
one-- and two--jet inclusive cross sections which have been obtained
with our method of slicing the phase space with invariant mass cuts.
In a short communication we used this method to calculate the
differential one-- and two--jet cross section 
as a function of $p_T$ integrated
over special rapidity intervals and compared it to recent experimental
data of the TOPAZ \cite{TOPAZ}  and AMY \cite{AMY} 
collaborations at TRISTAN. For this paper we
have calculated various one-- and two--jet distributions without
applying special cuts on kinematical variables of the initial or final
state dictated by the experimental analysis, although our approach is
particularly suitable for this.

The calculation of the cross sections proceeds as follows. For the DD and
DR components we use the phase space slicing method for the inclusive
one-- and two--jet cross sections. The details for the calculation of
the DD component below the cut are described in the previous section.
The DR component is identical with the work in \cite{KlaKra} for the direct
photoproduction of jets in low $Q^2$ $ep$ collisions. We must replace
the proton structure function by the photon structure function and
obtain the cross sections for the single--resolved contribution. With
the results presented in \cite{KlaKra} and in the previous section we are
able to calculate the inclusive cross section for one-- and two--jet
production. For the double--resolved contribution the NLO corrections
for the two--jet cross section are not available yet. Here only the LO
cross sections are at our disposal. To estimate the NLO corrections we
make use of the NLO cross sections for the inclusive one--jet case.
This cross section has the same structure as the NLO resolved
inclusive one--jet cross section for photoproduction $\g
p\to\mbox{jet} + X$, where the photon structure function at one vertex
is replaced by the proton structure function. For this cross section
the NLO corrections are known from earlier work \cite{KS}. We use
these results and transform the cross section to the $\g$ case by
replacing the proton structure function 
by the photon structure function. This
gives us the full NLO inclusive one--jet cross section for which we
present results later. We compare the NLO cross section with the RR LO
cross section and calculate the $k$ factor from it. Then the same $k$
factor is applied to the LO two--jet cross section. We checked this
procedure with the DD and DR cross sections, where we know also the
NLO two--jet cross sections. For these cases we found that the $k$
factors for the one-- and two--jet cross sections are approximately
equal. It is reasonable to expect this also to be the case for the RR
cross sections.

The further calculation of the DD and DR cross sections is based on
two separate contributions --- a set of two--body contributions and a
set of three--body contributions. Each set is completely finite, as all
singularities have been cancelled or absorbed into structure
functions. Each part depends separately on the cut--off $y$. If $y$ is
chosen large enough, the two parts determine physically well defined
two--jet and three--jet cross sections. However, our analytic
calculations are valid only for very small $y$, since terms $O(y)$
have been neglected in the analytic integrations. For very small $y$,
the two cross sections have no physical meaning. In this case, the
$(\ln y)$ terms force the two--body contributions to become negative,
whereas the three--body cross sections are large and positive. When
both contributions are added to yield a suitable inclusive cross
section, as for example the inclusive one--jet cross section, the
dependence on the cut--off $y$ will cancel. Then, the separation of
the two $y$ dependent contributions is only a technical device. The
cut--off only serves to distinguish the phase space regions, where the
integrations are done analytically, from those where they are
performed numerically. Furthermore, $y$ must be chosen sufficiently small
so that experimental cuts imposed on kinematical variables of the
final state do not interfere with the cancellation of the $y$
dependence. 

First we consider the inclusive one--jet cross section. We
choose the definition of the Snowmass meeting \cite{Sm} for combining
two nearly collinear partons. According to this definition, two
partons $i$ and $j$ are recombined if $R_{i,j}<R$ where
$R_i=\sqrt{(\eta_i-\eta_J)^2+(\phi_i-\phi_J)^2}$ is the distance
between parton $i$ and jet $J$ in the rapidity--azimuthal space.
$\eta_i$, $\phi_i$ and $\eta_J$, $\phi_J$ are the rapidities and the
azimuthal angles of parton $i$ and of the recombined jet $J$,
respectively. We choose $R=1$ in all the following results. The
Snowmass condition means that two partons are considered as two
separate jets or as a single jet depending whether they lie outside or
inside the cone with radius $R$ around the jet momentum. Unfortunately
this definition is not unique. In some cases it may happen that two
partons $i$ and $j$ qualify both as two individual jets $i$ and $j$
and as a recombined jet $ij$. In this case we count only the combined
jet following \cite{KuSo}. In NLO the final state may consist of two or
three jets. The three--jet sample consists of all three-body
contributions, which do not fulfill the cone condition. The inclusive
cross section will depend on the value of $R$ chosen. It will increase
with increasing $R$. 

Before we calculate the final results to be presented in the figures
we have made some checks of the NLO corrections to the one-- and 
two--jet cross sections. First we checked that the DD and DR cross
section are independent of the slicing cut $y$ if $y$ is chosen small
enough. This was the case for $y\le10^{-3}$ in all considered cases.
For $y>10^{-3}$ we observed some small $y$ dependence which is caused
by our approximation that we neglected contributions $O(y)$ in the 
analytical contributions to the two--jet cross section.
In the final evaluation we use $y=10^{-3}$. Furthermore we
tested that the sum of the NLO direct and the LO single resolved cross
section is independent of the factorization scale $M$. The same test
was performed for the sum of the NLO single resolved and the LO double
resolved cross section. These tests were done for the one-- and
two--jet cross sections separately. Similar checks have been reported
for the one--jet photoproduction cross section \cite{BKS}.

The input for our calculation is as follows. For the DR and RR cross
sections we need the parton distributions, $F_{i/\g}$, in the photon.
We have chosen the NLO set of Gl\"uck, Reya and Vogt (GRV) in the
$\overline{\mbox{MS}}$ scheme \cite{GRV}. This means that the DD, DR
and RR cross sections must also be calculated with the
$\overline{\mbox{MS}}$ subtraction. We choose all scales $\mu=M=p_T$
and calculate $\alpha_s(\mu)$ from the two--loop formula with $N_f=5$
massless flavours with $\Lambda_{\overline{\mbox{\tiny MS}}}^{(5)}=0.130\,
GeV$ equal to the $\Lambda$ value of the NLO GRV photon structure
function. The charm and bottom quarks are treated also as light
flavours with the boundary condition that the charm (bottom) contend
of the photon vanishes for $M^2\le m_c^2 (m_b^2)$ ($m_c=1.5\,GeV$,
$m_b= 5\,GeV$). We do not apply a special cut on the energy fractions
$x_a$ and $x_b$ in (\ref{f2}) but integrate from $x_{a,\min}$ and
$x_{b,\min}$ to $1$, where $x_{a,\min}$ and $x_{b,\min}$ are given by
kinematics (see (\ref{f5})).

In the following we show results for the three different c.m.\ energies: (i)
LEP1 with $\sqrt{S}=90\,GeV$ and $\theta_c=3^\circ$ in the photon
spectrum in (\ref{f1}), (ii) LEP2 with $\sqrt{S}=175\,GeV$ and
$\theta_c=30\,mrad$ and (iii) NLC in the TESLA design with
$\sqrt{S}=500\,GeV$ and the photon spectra given by the sum of the WWA
spectrum in (\ref{f1}) with $\theta_c=175\,mrad$ and the beamstrahlung
spectrum given in \cite{xx} with parameters
$\Upsilon_{\mbox{eff}}=0.039$ and $\sigma_z=0.5\,mm$ \cite{Sch}. (See
\cite{DreGod2} for the impact of various collider options.)

\begin{figure}[ht]
 \begin{center}
  \begin{picture}(15,10)
   \epsfig{file=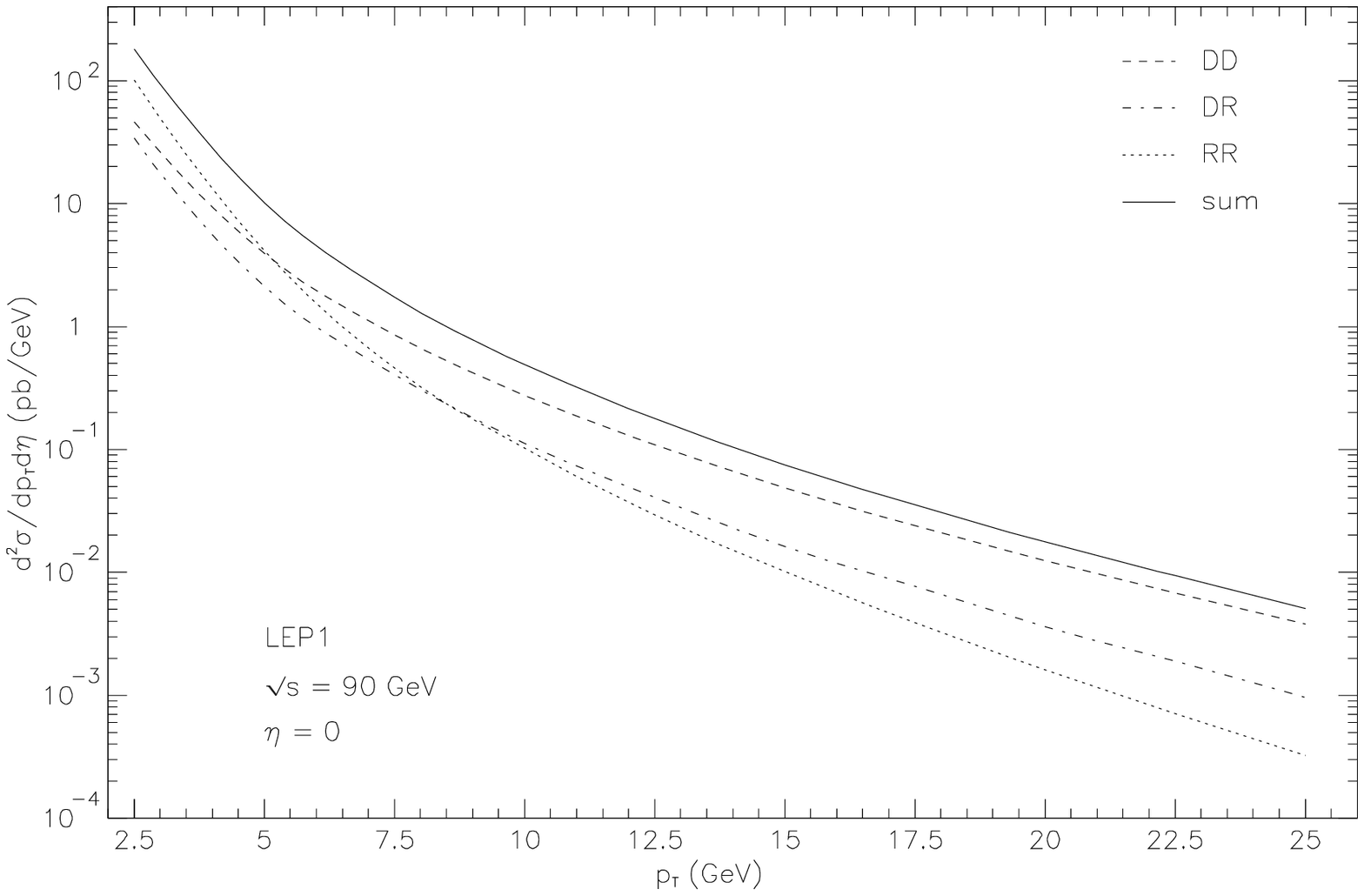,bbllx=25pt,bblly=240pt, bburx=530pt,bbury=575pt,%
           height=10cm}
  \end{picture}
  \caption{\label{bild1}}
 \end{center}
\end{figure}

\begin{figure}[ht]
 \begin{center}
  \begin{picture}(15,10)
   \epsfig{file=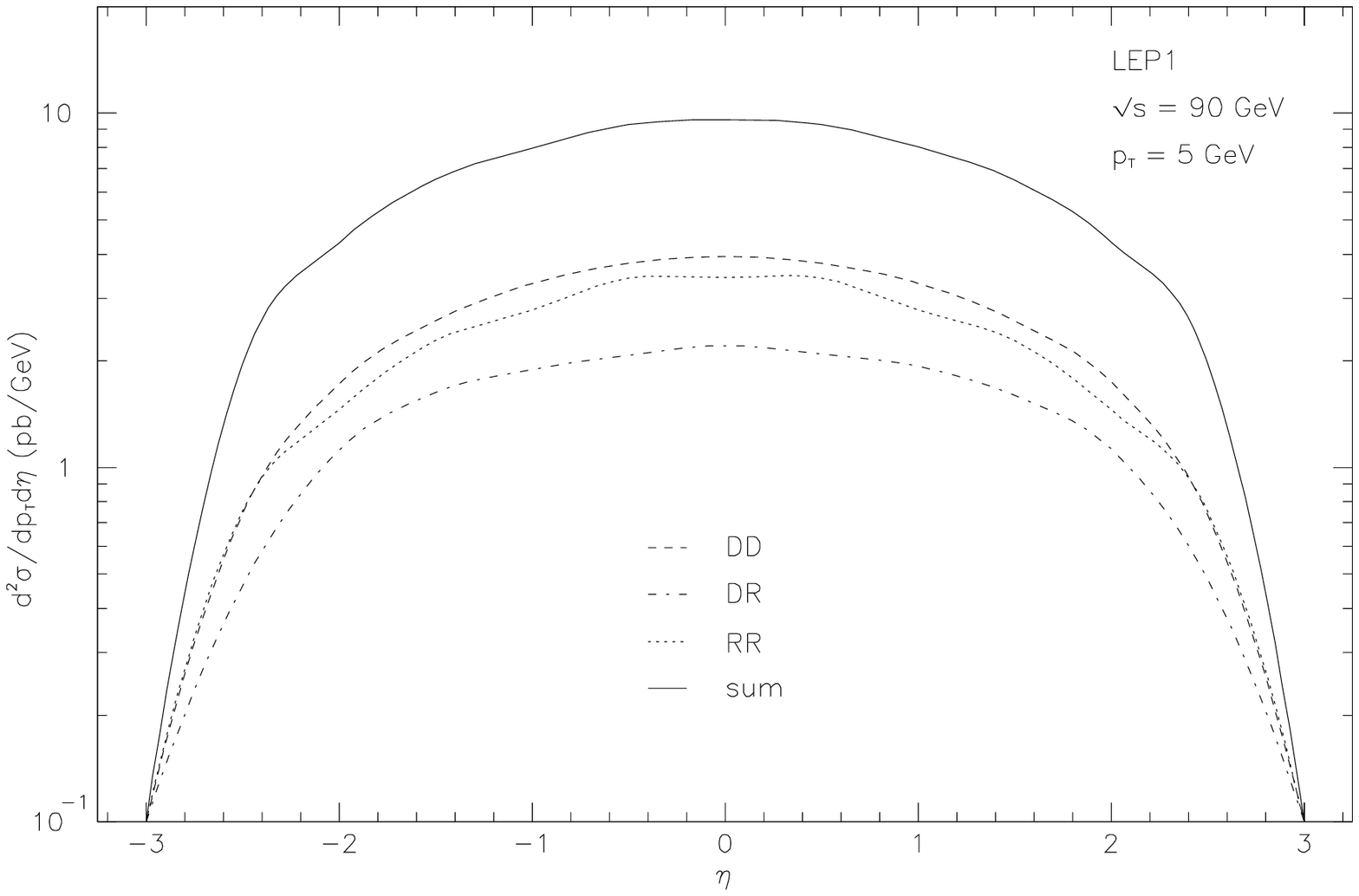,bbllx=25pt,bblly=240pt, bburx=530pt,bbury=575pt,%
           height=10cm}
  \end{picture}
  \caption{\label{bild2}}
 \end{center}
\end{figure}

\begin{figure}[ht]
 \begin{center}
  \begin{picture}(15,10)
   \epsfig{file=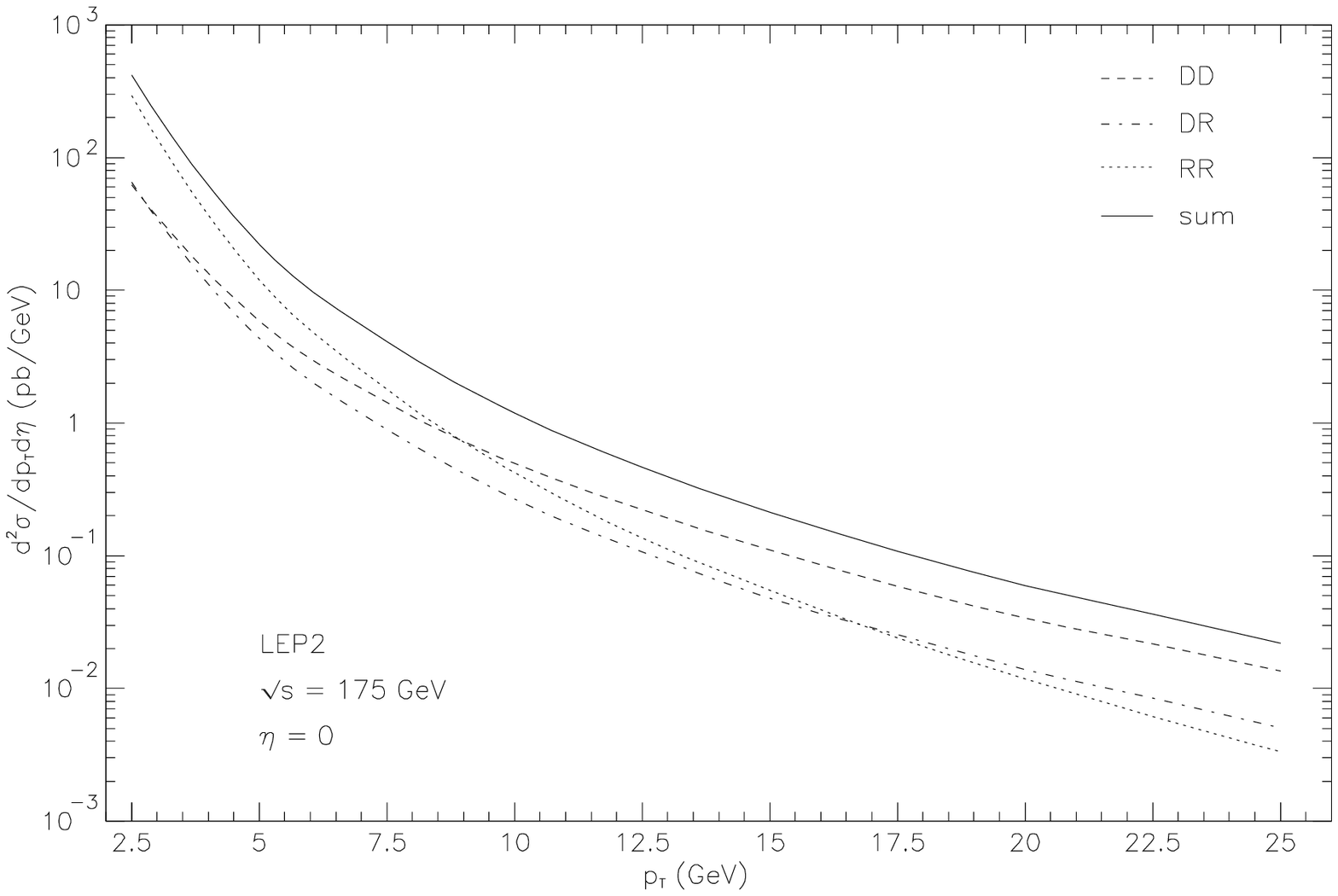,bbllx=25pt,bblly=240pt, bburx=530pt,bbury=575pt,%
           height=10cm}
  \end{picture}
   \caption{\label{bild3}}
 \end{center}
\end{figure}

\begin{figure}[ht]
 \begin{center}
  \begin{picture}(15,10)
   \epsfig{file=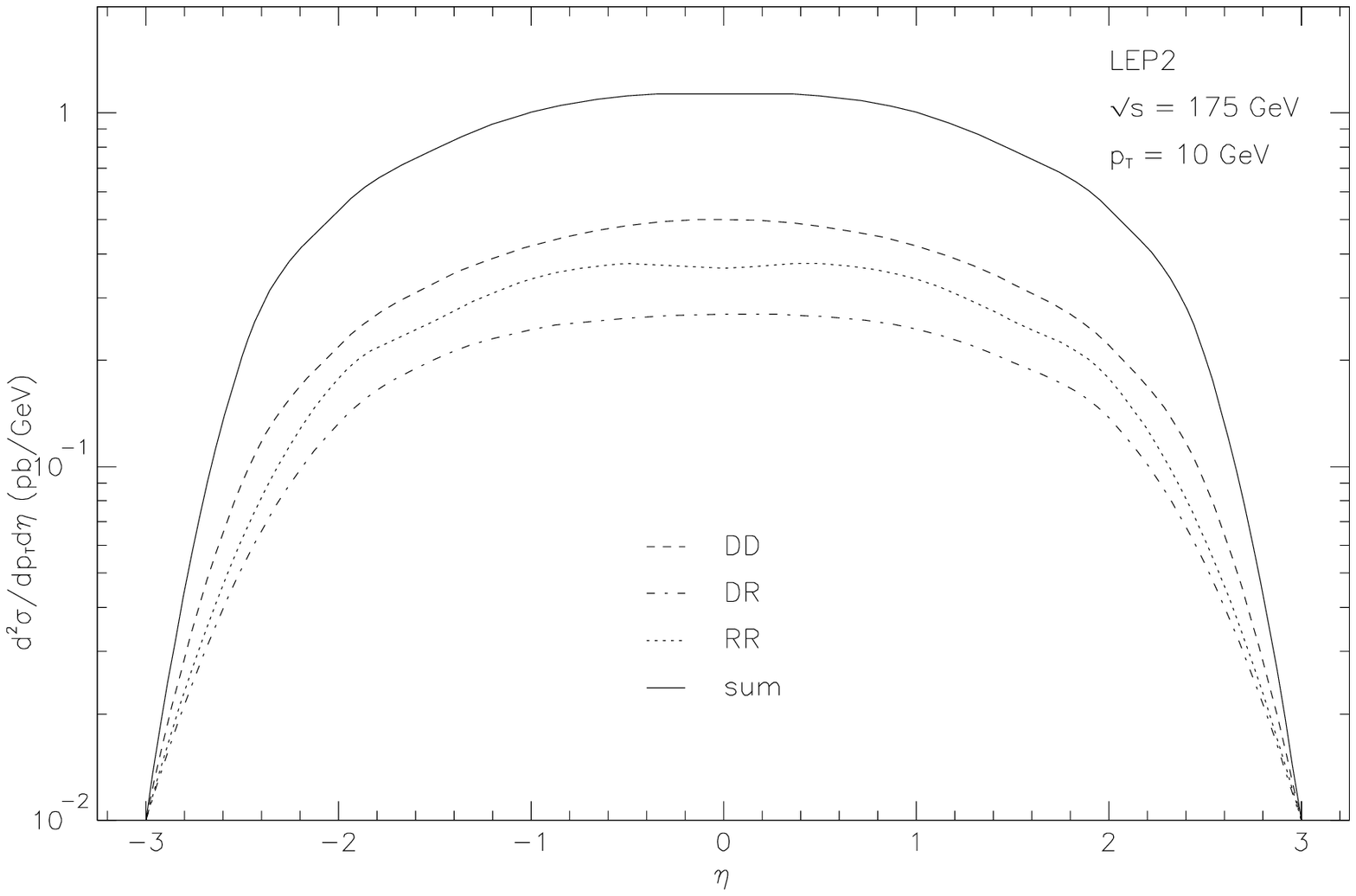,bbllx=25pt,bblly=240pt, bburx=530pt,bbury=575pt,%
           height=10cm}
  \end{picture}
  \caption{\label{bild4}}
 \end{center}
\end{figure}

\begin{figure}[ht]
 \begin{center}
  \begin{picture}(15,10)
   \epsfig{file=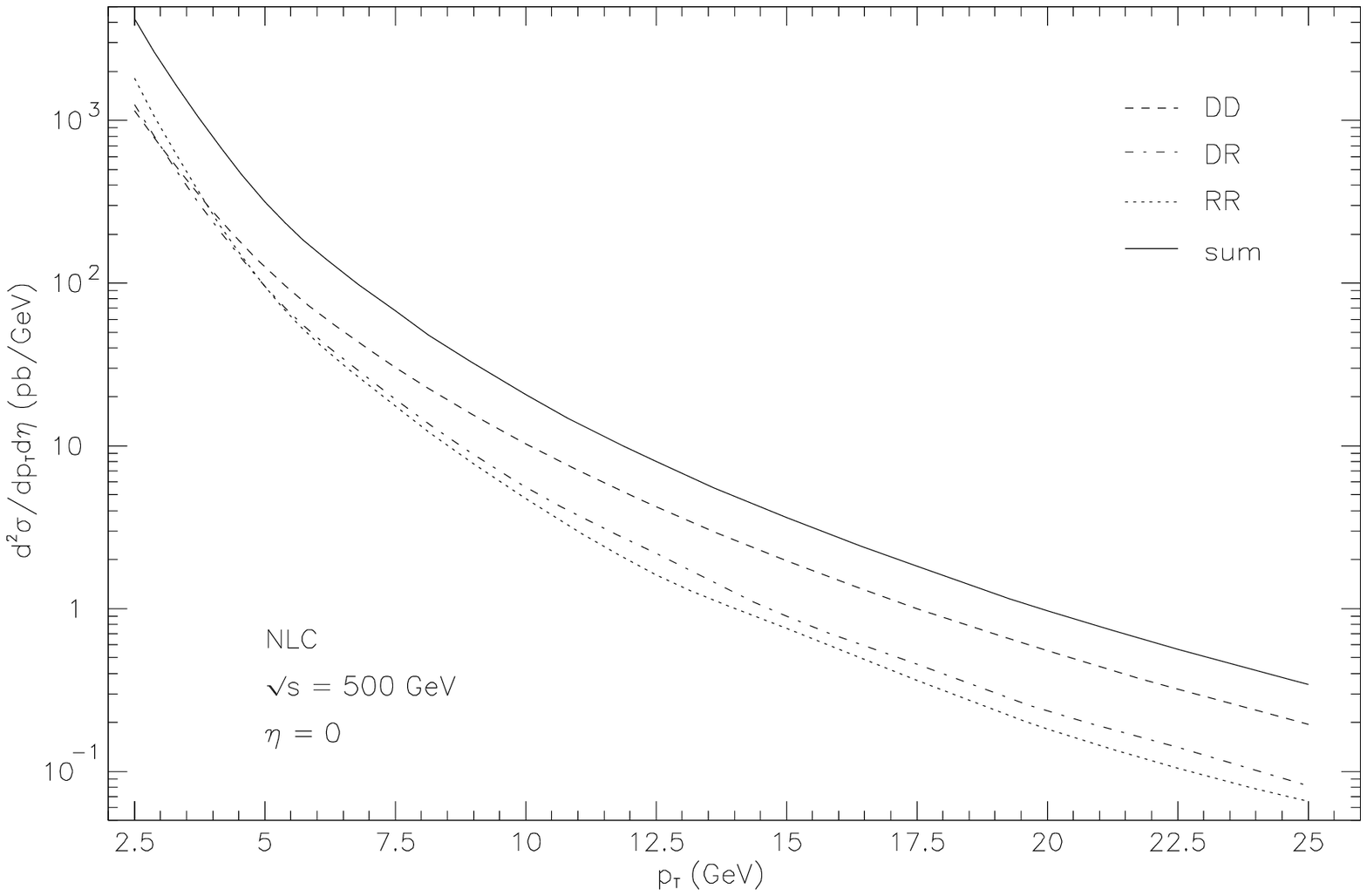,bbllx=25pt,bblly=240pt, bburx=530pt,bbury=575pt,%
           height=10cm}
  \end{picture}
  \caption{\label{bild5}}
 \end{center}
\end{figure}

\begin{figure}[ht]
 \begin{center}
  \begin{picture}(15,10)
   \epsfig{file=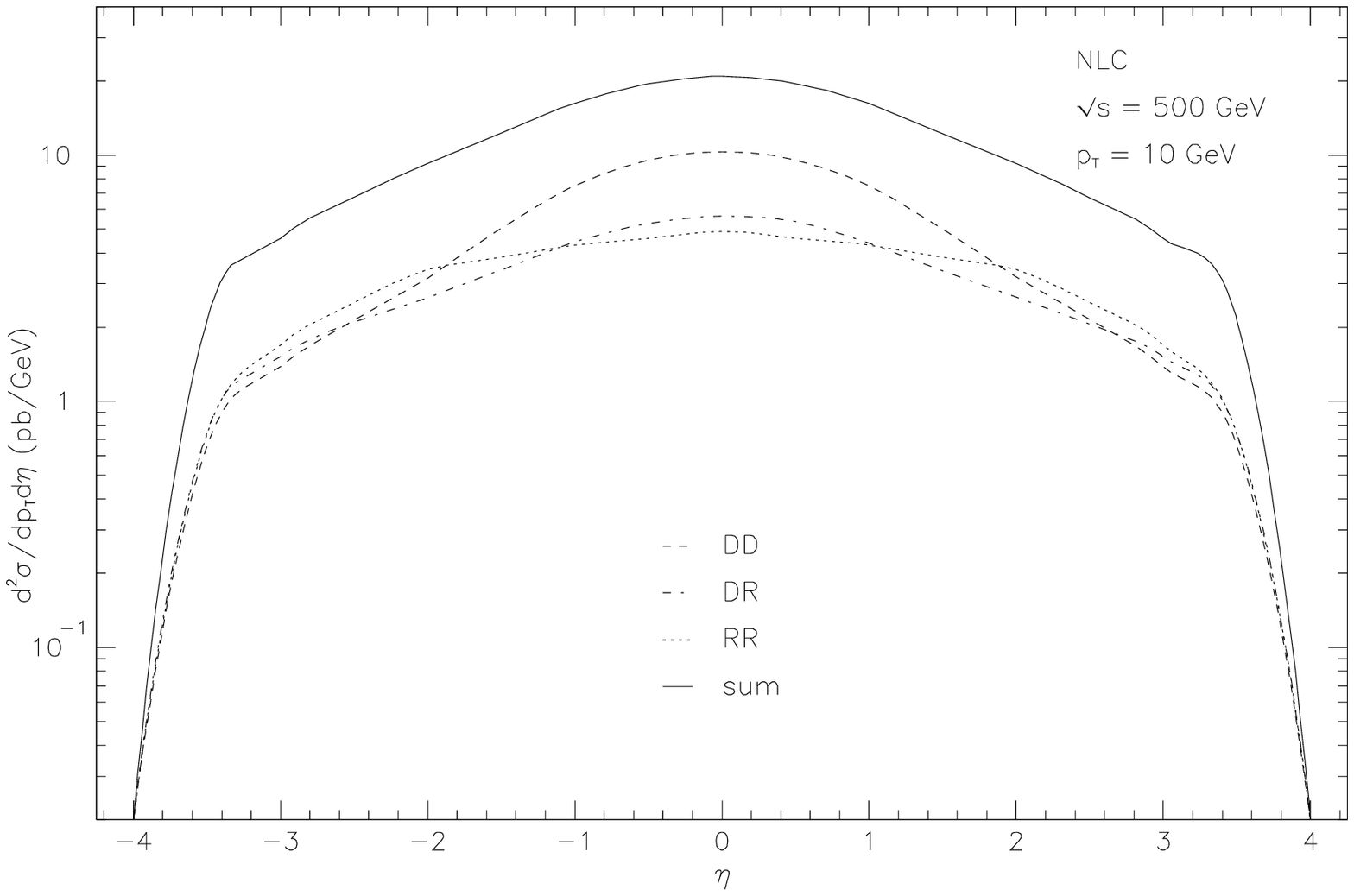,bbllx=25pt,bblly=240pt, bburx=530pt,bbury=575pt,%
           height=10cm}
  \end{picture}
  \caption{\label{bild6}}
 \end{center}
\end{figure}

First we show the inclusive one--jet cross sections $\d^2\sigma/\d
p_T\d\eta$ for the three machines. In Fig.~\ref{bild1} this cross section is
plotted as a function of $p_T$ for rapidity $\eta=0$, where this cross
section is maximal. Only the NLO predictions are plotted for
the DD, DR and RR cross sections and for the sum of all contributions.
Below $p_T=5\,GeV$ the RR component is dominant whereas for the larger
$p_T$ the DD cross section gives the largest contribution. At
$p_T=5\,GeV$ the DD and RR cross sections are equal. Above $p_T=10\,GeV$
the DR cross section is larger than the RR cross section. The rapidity
distribution at $p_T=5\,GeV$ is shown in Fig.~\ref{bild2}, again for the three
components and for the sum. The slight variations in the RR curve are
caused by the limited accuracy of the numerical integrations. 
All three cross sections must be symmetric
at $\eta=0$, where they exhibit a rather broad plateau. The same
distributions, i.e.~$\d^2\sigma/\d p_T\d\eta$ for $\eta=0$ and
$\d^2\sigma/\d p_T\d\eta$ for $p_T=10\,GeV$ are shown in 
Fig.~\ref{bild3} and \ref{bild4} for LEP2.
The qualitative behaviour of the one--jet cross sections is similar.
Due to the larger c.m.\ energy the cross sections for LEP2 are larger
than for LEP1. The RR contribution is more significant now. The DD 
distribution crosses the RR distribution at larger $p_T$ than in the
LEP1 case. The rapidity distribution at $p_T=10\,GeV$ looks very similar
to that in Fig.~\ref{bild2}. At this $p_T$ value all three components make
significant contributions to the total sum. In the NLC case the
pattern is somewhat different as can be seen in 
Fig.~\ref{bild5} and \ref{bild6}. Due to the further increase of the
c.m.\ energy the cross section is increased by an order of
magnitude as compared to the LEP2 cross section. The hierarchy between
the DD, DR and RR components has changed somewhat.
The RR cross sections lie for $p_T>5\,GeV$ below the DD and DR cross sections.
All three components cross each other below $p_T=5\,GeV$, where RR starts
to dominate. Above $p_T=5\,GeV$
the $p_T$ distribution is dominated by the DD component. The total
rapidity distribution at $p_T=10\,GeV$ is less flat around $\eta=0$.
This originates from the DD and DR part. The DD part has the steepest
behaviour towards the kinematic boundaries. The different functional
behaviour of $\d\sigma/\d p_T$ and $\d\sigma/\d\eta$ in 
Fig.~\ref{bild5} and \ref{bild6}
as compared to the cross sections for LEP2 comes from the changed
photon spectra which is now a superposition of the WWA and the
beamstrahlung spectrum. The rapidity distribution for the NLC looks
similar to the results obtained recently for charm quark production in
two--photon collisions \cite{CGKKKS}. However, the difference between the LEP2
and NLC rapidity distributions is even more drastic for heavy quark
production than in Fig.~\ref{bild4} and \ref{bild6}. 
By comparing with the LO one--jet
cross section, where the same photon structure function and the same
$\alpha_s$ as in the NLO cross section is used we obtained the $k$ factors
for the RR contributions. They are $k=1.85$, $1.90$ and $1.90$ for the
three cases LEP1, LEP2 and NLC. We shall use these $k$ factors for
correcting the LO RR predictions of the two--jet cross sections.

Next we consider the results on inclusive two--jet production
which will be shown in the following figures for the LEP2 case. In
Fig.~\ref{bild7}, \ref{bild8} and \ref{bild9}, 
we present $\d^3\sigma/\d p_T\d\eta_1\d\eta_2$ as a
function of $p_{T_1}$ for $\eta_1=0$ and various choices of
$\eta_2=0$, $1$, $2$. Here, $p_{T_1}$ and $\eta_1$ are the transverse
momentum and the rapidity of the so--called trigger jet. $\eta_2$ is
the rapidity of the second jet, so that $p_{T_1}$ and $p_{T_2}$ are
the two highest transverse momenta of 
the three--jet configuration. For exactly two
jets in the final state, we have $p_{T_1}=p_{T_2}$. 
In Fig.~\ref{bild8} and \ref{bild9},
we can see how the cross section decreases when $\eta_2$ is chosen
away from the maximum region at $\eta_2=0$. In particular the RR component
becomes more important for increasing $p_T$ when $\eta_2$ increases.
$\eta_1=0$ is always kept fixed. Comparing the cross sections in
Fig.~\ref{bild7} and \ref{bild8} 
 we see that the sum of the three components hardly
changes when $\eta_2=1$ instead of $\eta_2=0$, although the relation of
the DD, DR and RR cross sections is different in the two cases. 
We have studied the inclusive two--jet cross section also as a
function of $\eta_1$ and $\eta_2$ for fixed $p_{T_1}$. As an example, 
we show for LEP2 the two--dimensional distribution $\d^3\sigma/\d
p_T\d\eta_1\d\eta_2$ for $p_{T_1}=10\,GeV$ in form of a lego--plot in the
intervals $\eta_1$, $\eta_2\in[-3.0,3.0]$. The cross sections for the
DD, DR and RR contributions and for the sum are shown separately. The
DR cross section contains both contributions with the resolved photon
at the upper or the lower vertex. If they are considered separately we
checked that they are symmetric for $\eta_1\leftrightarrow\eta_2$. 

For the NLC case we present only $\d^3\sigma/\d p_T\d\eta_1\d\eta_2$ as
a function of $p_{T_1}$ for $\eta_1=\eta_2=0$ where the two--jet cross
section is maximal. This is shown in Fig.~\ref{bild11}. Comparing with the
corresponding cross section for LEP2 in Fig.~\ref{bild7} we notice the increase
of the cross section by more than a factor 10 and that the RR component
is somewhat more significant also for larger $p_T$. The cross section
$\d^3\sigma/\d p_T\d\eta_1\d\eta_2$ as a function of $\eta_2$ for
$p_{T_1}=10\,GeV$ and $\eta_1=0$ is presented in 
Fig.~\ref{bild12} for the three
components DD, DR, RR and the sum. Only the RR cross section is rather
flat near the maximum $\eta_2=0$. The other components are much
steeper away from $\eta_2=0$ as it is the case also in the sum
similar to the one--jet cross section shown in Fig.~\ref{bild6}. 

It is clear that many more distributions or partially integrated cross
sections using other two--jet variables can be calculated with the
phase space slicing method. So, for example, one could study kinematic
regions where either the DD or the RR cross section is enhanced, as
has been done for two--jet production in $\g p$ processes \cite{KlaKra} by
making cuts in $x_{\g}$, the fraction of the photon energy
participating in the hard scattering process. Other interesting topics
are the cone dependence of the inclusive cross sections or the study
of the invariant mass distribution of the two jets for different
rapidities or the angular distribution of the two jets to test the
parton--parton scattering dynamics in a different way.

\section{Summary and Conclusions}
Various inclusive one-- and two--jet cross sections have been
calculated for the direct, single resolved and double resolved
contributions as a function of $p_T$ and jet rapidities in NLO. For
the double resolved two--jet cross section the NLO corrections are
estimated with a $k$ factor taken from the inclusive one--jet cross
section. For the direct and single resolved components all NLO
corrections are fully evaluated. Infrared and collinear singularities
are cancelled with the phase space slicing method using an invariant
mass cut--off. This method is particularly useful for incorporating
cuts on the final state and for obtaining results with different
choices of jet algorithms. Analytical formulas for the different
contributions giving the dependence on the slicing parameter are
derived for the direct contribution. The same results for the single
resolved contribution are taken from the corresponding calculation of
jet production in $\g p$ processes \cite{KlaKra}.

Numerical results for the inclusive one--jet and two--jet 
cross sections in three energy ranges corresponding to LEP1, LEP2 and
NLC have been presented. In the NLC case we have calculated the photon 
spectra from a superimposition of bremsstrahlung and beamstrahlung
spectra. For this case the photon spectra lead to quite different
rapidity distributions as compared to LEP1 and LEP2 where we have only
the bremsstrahlung spectrum. The hierarchy of direct, single and double
resolved cross section is very similar for the three machines. The RR
component dominates only for rather small $p_T\le5\,GeV$ in all cases.
For larger $p_T$ the direct contribution is dominant. 

Since our results have been successfully  tested already at smaller energies
by comparing with one-- and two--jet measurements of the TOPAZ and AMY
collaborations we are confident that our results are also reliable for
the larger energies achievable at LEP2 and NLC. 

\newpage

\begin{appendix}
\section{Virtual Correction}
\be
   V_\g(s,t,u)&=& \left[-\frac{2}{\eps^2}-\frac{3}{\eps}+\frac{2\pi^2}{3}
   -7+\ln^2\frac{-t}{s}+\ln^2\frac{-u}{s}\right]T_\g^{(\eps)}(s,t,u)\nn\\
   & & +2\ln\frac{-t}{s}+2\ln\frac{-u}{s}+3\frac{u}{t}\ln\frac{-t}{s}
       +3\frac{t}{u}\ln\frac{-u}{s}\nn\\
   & & +\left(2+\frac{u}{t}\right)\ln^2\frac{-u}{s}
       +\left(2+\frac{t}{u}\right)\ln^2\frac{-t}{s}\label{a1}
\ee

\section{Final State Correction}
\be
   F_\g(s,t,u) &=& \left[\frac{2}{\eps^2}+\frac{3}{\eps}+7-2\ln^2y_F-3\ln
   y_F\right]T_\g^{(\eps)}(s,t,u)
\ee 

\section{Initial State Correction}
\be
   I_\g(z;s,t,u) &=& \left[-\frac{1}{\eps}\frac{1}{N_CQ^2}
   P_{q\leftarrow\g}(z)
   +1+\ln\left(\frac{y_I(1-z)}{z}\right)\left(2z^2-2z+1\right)\right]
   T_q^{(\eps)}(s,t,u)
\ee
Here one needs the Altarelli--Parisi splitting function for the
process photon $\to$ quark, which is
\be
   P_{q\leftarrow\g}(z)&=& N_CQ^2\left[2z^2+2z+1\right]
\ee

\end{appendix}

\newpage
\section{Figure Captions}
\begin{enumerate}
\item The three body diagrams for $\g\g\to q\bar{q}g$.

\item Three body diagrams with final and initial state singularities.

\item Kinematic diagram for the three--body final state defining the 
      angles in the c.m.~system of partons $p_1$ and $p_2$. 

\item NLO inclusive one--jet cross section $\d^2\sigma/\d p_T\d\eta$
      as a function of $p_T$ at $\eta=0$ for direct (DD), single
      resolved (DR) and double resolved (RR) contributions and for the
      sum of all components. LEP1 photon spectra, $\sqrt{S}=90\,GeV$.

\item NLO inclusive one--jet cross section as a function of $\eta$ at 
      $p_T=5\,GeV$ and for DD, DR, RR and the sum. LEP1 photon spectra,
      $\sqrt{S}=90\,GeV$.

\item Same as Fig.~4. LEP2 photon spectra, $\sqrt{S}=175\,GeV$.

\item Same as Fig.~5, $p_T=10\,GeV$. LEP2 photon spectra, $\sqrt{S}=175\,GeV$

\item Same as Fig.~4. NLC--TESLA photon spectra. $\sqrt{S}=500\,GeV$.

\item Same as Fig.~5, $p_T=10\,GeV$. NLC--TESLA photon spectra. 
      $\sqrt{S}=500\,GeV$.

\item NLO inclusive two--jet cross section as a function of $p_T$ for
      $\eta_1=\eta_2=0$ for direct (DD), single resolved (DR), double
      resolved (RR) and sum. LEP2 photon spectra. $\sqrt{S}=175\,GeV$.
      RR component estimated with $k$ factor.

\item Same as Fig.~10 for $\eta_1=0$, $\eta_2=1$.

\item Same as Fig.~10 for $\eta_1=0$, $\eta_2=2$.

\item NLO triple differential cross section $\d^3\sigma/\d
      p_T\d\eta_1\d\eta_2$ for $p_T=10\,GeV$ as a function of $\eta_1$
      and $\eta_2$ for DD, DR, RR components and the sum.
      LEP2 photon spectra. $\sqrt{S}=175\,GeV$.

\item Same as Fig.~10. NLC--TESLA photon spectra. $\sqrt{S}=500\,GeV$.

\item NLO inclusive two--jet cross section as a function of $\eta_2$
      for $p_T=10\,GeV$ and $\eta_1=0$. NLC--TESLA photon spectra.
      $\sqrt{S}=500\,GeV$.

\end{enumerate}

\newpage

\begin{figure}[ht]
 \begin{center}
  \begin{picture}(15,10)
   \epsfig{file=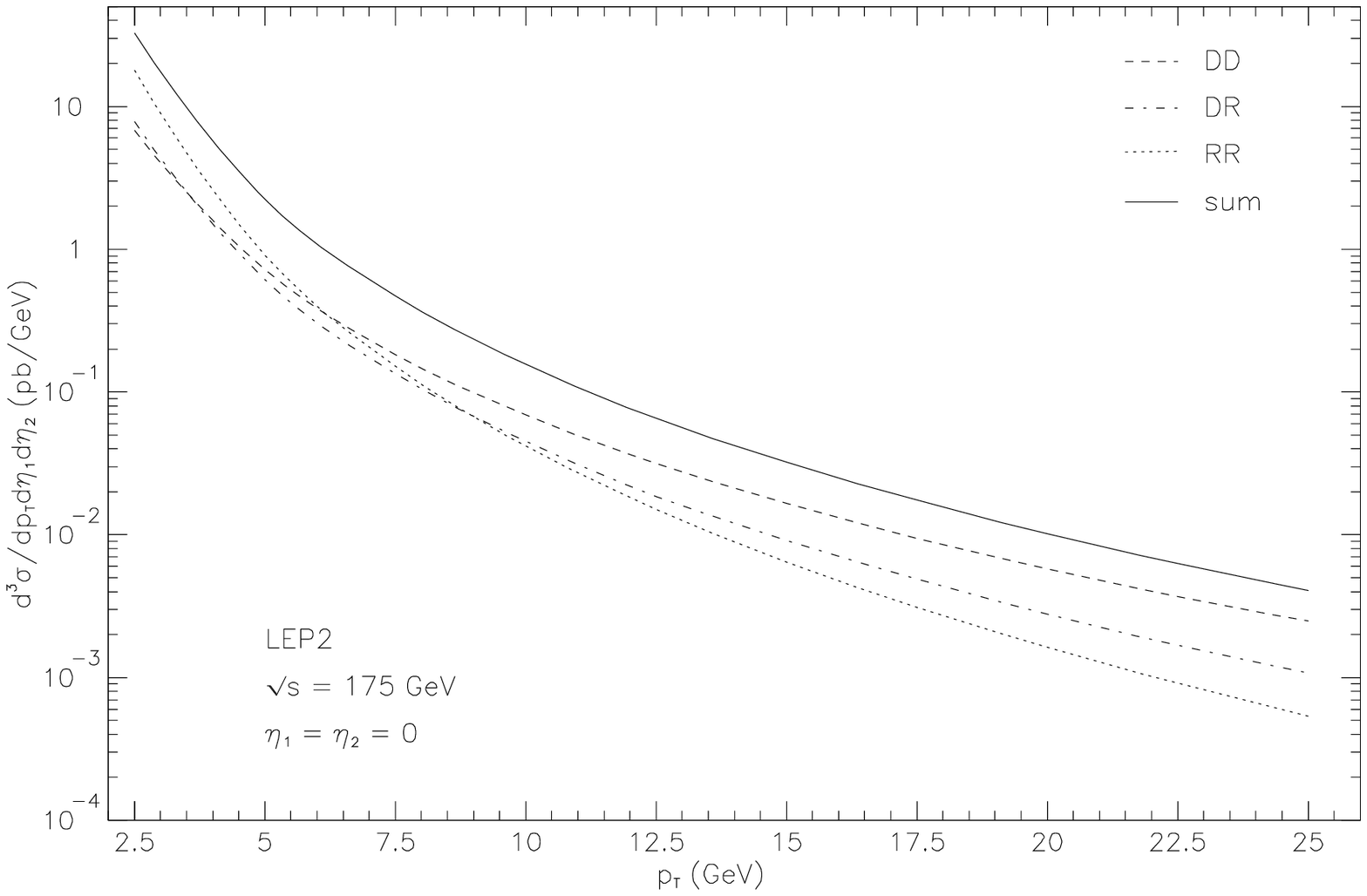,bbllx=25pt,bblly=240pt, bburx=530pt,bbury=575pt,%
           height=10cm}
  \end{picture}
  \caption{\label{bild7}}
 \end{center}
\end{figure}

\begin{figure}[ht]
 \begin{center}
  \begin{picture}(15,10)
   \epsfig{file=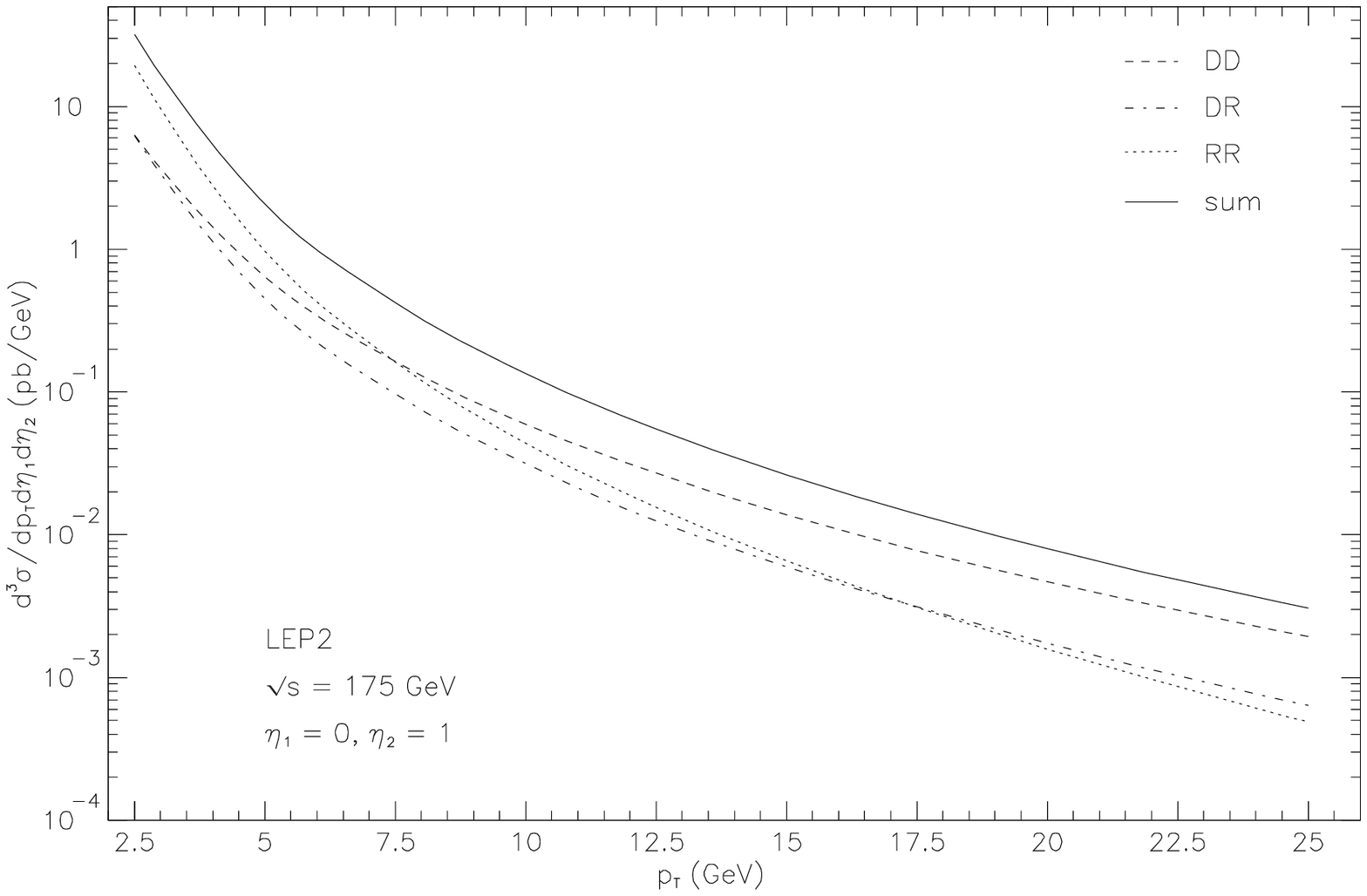,bbllx=25pt,bblly=240pt, bburx=530pt,bbury=575pt,%
           height=10cm}
  \end{picture}
  \caption{\label{bild8}}
 \end{center}
\end{figure}

\begin{figure}[ht]
 \begin{center}
  \begin{picture}(15,10)
   \epsfig{file=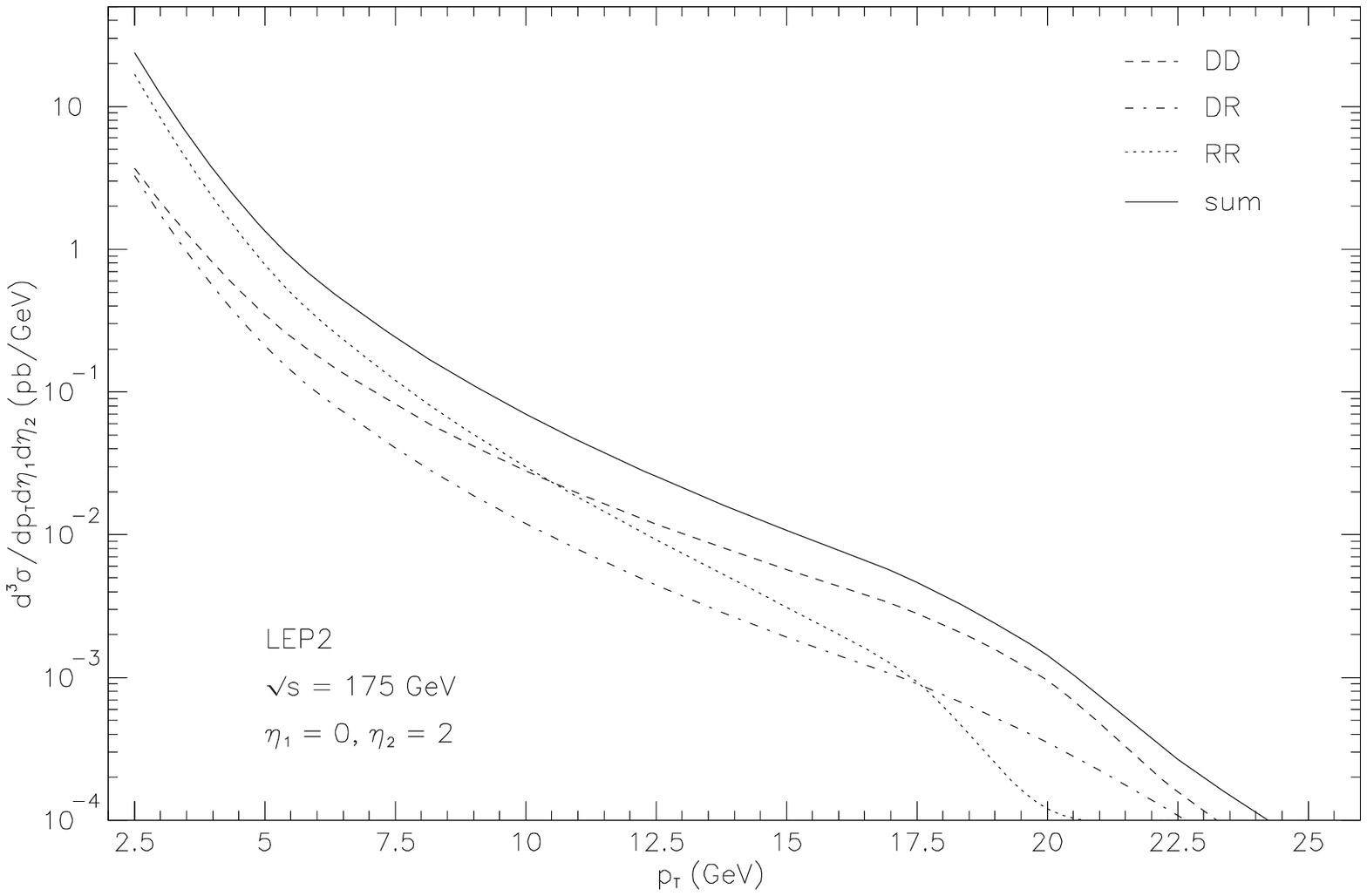,bbllx=25pt,bblly=240pt, bburx=530pt,bbury=575pt,%
           height=10cm}
  \end{picture}
  \caption{\label{bild9}}
 \end{center}
\end{figure}

\begin{figure}[ht]
 \begin{center}
  \begin{picture}(15,15)
   \epsfig{file=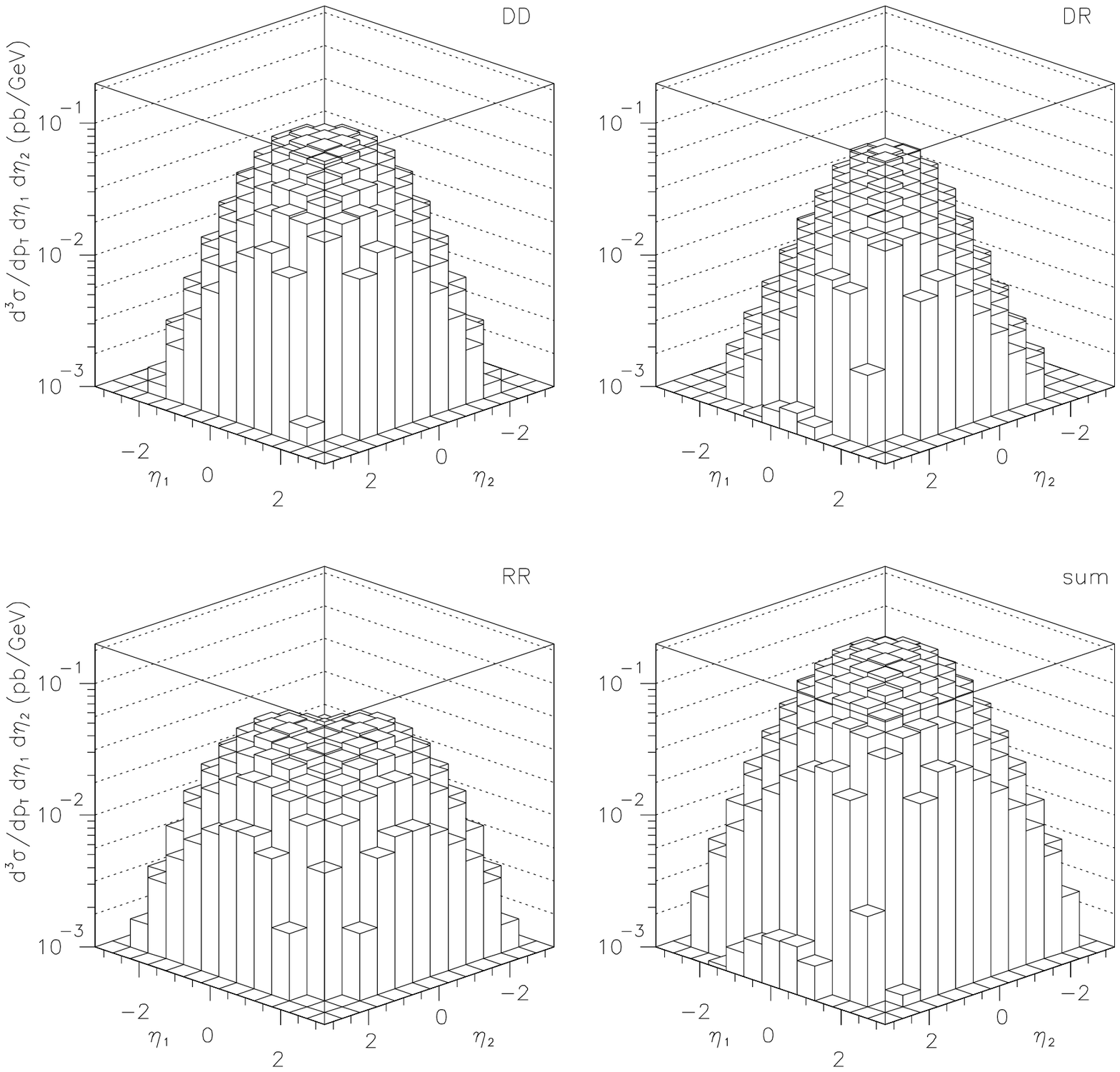,bbllx=20pt,bblly=165pt, bburx=530pt,bbury=655pt,%
           height=15cm}
  \end{picture}
  \caption{\label{bild10}}
 \end{center}
\end{figure}

\begin{figure}[ht]
 \begin{center}
  \begin{picture}(15,10)
   \epsfig{file=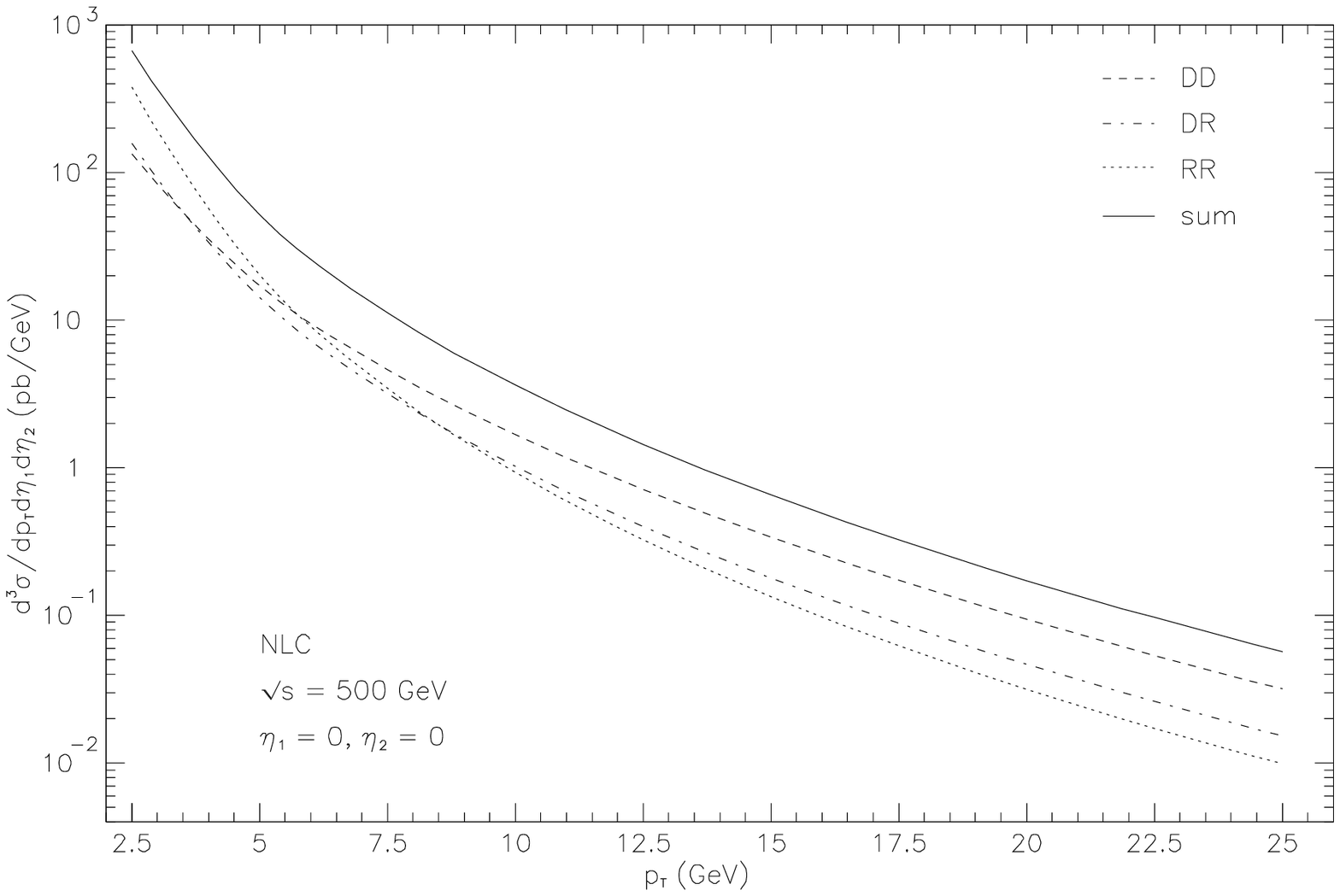,bbllx=25pt,bblly=240pt, bburx=530pt,bbury=575pt,%
           height=10cm}
  \end{picture}
  \caption{\label{bild11}}
 \end{center}
\end{figure}

\begin{figure}[ht]
 \begin{center}
  \begin{picture}(15,10)
   \epsfig{file=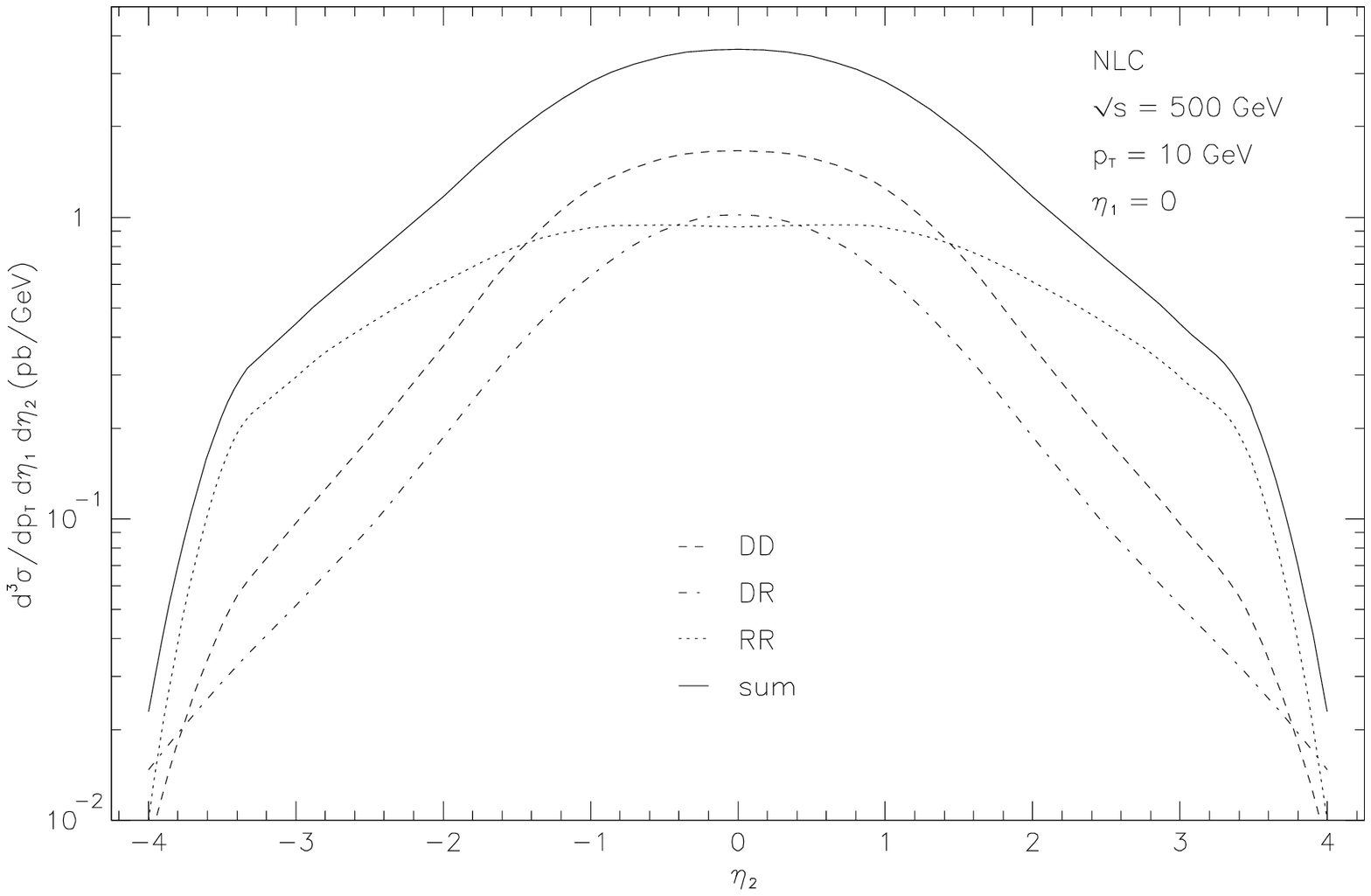,bbllx=25pt,bblly=240pt, bburx=530pt,bbury=575pt,%
           height=10cm}
  \end{picture}
  \caption{\label{bild12}}
 \end{center}
\end{figure}

\end{document}